%% file: mgirardi.tex
\documentclass{aa}  
\usepackage{graphicx}
\usepackage{txfonts}
\newcommand{\mincir}{\raise -2.truept\hbox{\rlap{\hbox{$\sim$}}\raise5.truept
\hbox{$<$}\ }}
\newcommand{\magcir}{\raise -2.truept\hbox{\rlap{\hbox{$\sim$}}\raise5.truept
\hbox{$>$}\ }}
\newcommand{\siml}{\raise -2.truept\hbox{\rlap{\hbox{$\sim$}}\raise5.truept
\hbox{$<$}\ }}
\newcommand{\simg}{\raise -2.truept\hbox{\rlap{\hbox{$\sim$}}\raise5.truept
\hbox{$>$}\ }}
\newcommand{\be}{\begin{equation}}
\newcommand{\ee}{\end{equation}}
\newcommand{\ba}{\begin{eqnarray}}
\newcommand{\ea}{\end{eqnarray}}
\newcommand {\kpc} {$h_{70}^{-1}$ kpc $\;$}
\newcommand {\kpcc} {$h_{70}^{-1}$ kpc}

\newcommand {\h} {$h_{70}^{-1}$ Mpc$\;$}
\newcommand {\hh} {$h_{70}^{-1}$ Mpc}
\newcommand {\hhh} {\;h_{70}^{-1} \mathrm{Mpc}}
\newcommand {\ks} {km~s$^{-1} \;$}
\newcommand {\kss} {km~s$^{-1}$}

\newcommand {\mqua} {$\times 10^{14}\;h_{70}^{-1}\;M_{\odot} \;$}

\newcommand {\mqui} {$\times 10^{15}\;h_{70}^{-1}\;M_{\odot} \;$}
\newcommand {\mquii} {$\times 10^{15}\;h_{70}^{-1}\;M_{\odot}$}
\newcommand {\ml} {$h_{70}\;M_{\odot}/L_{\odot} \;$}

\newcommand{\degree}{\ensuremath{\mathrm{^\circ}}}
\newcommand{\arcm}{\ensuremath{\mathrm{^\prime}\;}}
\newcommand{\arcs}{\ensuremath{\arcmm\hskip -0.1em\arcmm \;}}
\newcommand{\arcmm}{\ensuremath{\mathrm{^\prime}}}
\newcommand{\arcss}{\ensuremath{\arcmm\hskip -0.1em\arcmm}}

\newcommand{\dotsec}{\,\rlap{\hbox{$\mathrm{^s}$}}{\hbox{$.$}}\,}
\newcommand{\xmm}{{\it XMM-Newton}}

\begin{document}
   \title{Internal dynamics of Abell 2254: a merging galaxy cluster
with a clumpy, diffuse radio emission}  

   \author{M. Girardi\inst{1,2}
          \and
S. Bardelli\inst{3}
          \and
R. Barrena\inst{4,5}
          \and
W. Boschin\inst{6}
          \and
F. Gastaldello\inst{7,8}
          \and
M. Nonino\inst{2}
}

   \offprints{M. Girardi, \email{mgirardi@oats.inaf.it}}

   \institute{ 
     Dipartimento di Fisica dell'Universit\`a degli Studi
     di Trieste - Sezione di Astronomia, via Tiepolo 11, I-34143
     Trieste, Italy\\ 
\and INAF - Osservatorio Astronomico di Trieste,
     via Tiepolo 11, I-34143 Trieste, Italy\\ 
\and INAF - Osservatorio Astronomico di Bologna, via Ranzani 1, 
I-40127, Bologna, Italy\\
\and Instituto de
     Astrof\'{\i}sica de Canarias, C/V\'{\i}a L\'actea s/n, E-38205 La
     Laguna (Tenerife), Canary Islands, Spain\\ 
\and Departamento de
     Astrof\'{\i}sica, Universidad de La Laguna, Av. del
     Astrof\'{\i}sico Franciso S\'anchez s/n, E-38205 La Laguna
     (Tenerife), Canary Islands, Spain\\
\and 
     Fundaci\'on Galileo
     Galilei - INAF, Rambla Jos\'e Ana Fern\'andez Perez 7, E-38712
     Bre\~na Baja (La Palma), Canary Islands, Spain\\ 
\and INAF-IASF Milano, Via Bassini 15, I-20133 Milano, Italy\\
\and 
Department of Physics and Astronomy, University of California at Irvine, 4129, 
Frederick Reines Hall, Irvine, CA, 92697-4575, USA
}

\date{Received  / Accepted }

\abstract{The mechanisms giving rise to diffuse radio emission in
  galaxy clusters, and in particular their connection with cluster
  mergers, are still debated.}{We explore the internal dynamics of
  Abell 2254, which has been shown to host a very clumpy and irregular
  radio halo.}  {Our analysis is mainly based on redshift data for 128
  galaxies acquired at the Telescopio Nazionale Galileo. We combine
  galaxy velocities and positions to select 110 cluster galaxies and
  analyze its internal dynamics.  We also use new ($g^{\prime}$,
  $r^{\prime}$, $i^{\prime}$) photometric data acquired at the Isaac
  Newton Telescope and (V,$i^{\prime}$) photometric data available in
  the Subaru Archive. X-ray data from the XMM-Newton Science Archive
  are analyzed to study the hot gas component.}{We estimate the
  cluster redshift $\left<z\right>=0.177$, a high line-of-sight
  (LOS) velocity dispersion, $\sigma_{\rm V}\sim 1350$ \ks, and the
  X-ray temperature $kT\sim \,$6.4 keV.  Both our optical and
  X-ray analyses reveal a complex dynamical activity.  The analysis
  of the 2D galaxy distribution reveals the presence of two density
  peaks, one at the East and the other at the West. Using the full 3D
  information we detect a high velocity ($\Delta V_{\rm rf,LOS}\sim
  3000$ \kss), low mass ($\sigma_{\rm V}\sim 200$--500 \kss) group at
  the position of the 2D eastern peak. For the main system we compute
  a velocity dispersion $\sigma_{\rm V}\sim 1000$--1200 \kss.  In the
  assumption of a bimodal system we estimate a mass $M_{\rm
    sys}=1.5$--2.9 \mquii. The X-ray morphological analysis, based on
  power ratios, centroid shifts, and concentration parameter, confirms
  that Abell 2254 is a dynamically disturbed cluster. The X-ray
  isophotes are elongated toward the eastern direction, in agreement
  with a merger in the post core-crossing phase.  A simple bimodal
  model finds that data are consistent with a bound, outgoing
  subcluster observed a few fractions of Gyr after the core
  crossing. However, both optical and X-ray analyses suggest that the
  main system is, at its time, a non relaxed structure, indicating N-S
  as a possible direction for a past accretion.  } {We conclude that
  Abell 2254, for its mass and merging structure, fits well among
  typical clusters with radio halos.  We shortly discuss as the
  particular irregularity of the radio halo might be linked to the
  complexity of the Abell 2254 structure.}

   \titlerunning{Internal dynamics of Abell 2254}
   \maketitle
%

\section{Introduction}
\label{intr}

Merging processes constitute an essential ingredient of the evolution
of galaxy clusters (see Feretti et al. \cite{fer02b} for a review). An
interesting aspect of these phenomena is the possible connection
between cluster mergers and extended, diffuse radio sources: halos and
relics. The synchrotron radio emission of these sources demonstrates
the existence of large-scale cluster magnetic fields and of
widespread relativistic particles. Cluster mergers have been proposed
to provide the large amount of energy necessary for electron
re-acceleration to relativistic energies and for magnetic field
amplification (Tribble \cite{tri93}; Feretti \cite{fer99}; Feretti
\cite{fer02a}; Sarazin \cite{sar02}). Radio relics (``radio gischts''
as referred to by Kempner et al. \cite{kem04}), which are polarized
and elongated radio sources located in the cluster peripheral regions,
seem to be directly associated with merger shocks (e.g., Ensslin et
al. \cite{ens98}; Roettiger et al. \cite{roe99}; Ensslin \&
Gopal-Krishna \cite{ens01}; Hoeft et al. \cite{hoe04}). Radio halos
are unpolarized sources that permeate the cluster volume in a similar
way to the X-ray emitting gas of the intracluster medium (hereafter
ICM). Radio halos are more likely to be associated with the turbulence
following a cluster merger, although the precise radio formation
scenario remains unclear (re-acceleration vs. hadronic models e.g.,
Brunetti et al.  \cite{bru09}; Ensslin et al.  \cite{ens11}).  Recent
semi-analytical calculations in the framework of the turbulent
re-acceleration scenario have allowed to derive the expectations for
the statistical properties of giant radio halos, in agreement with
present observation that halos are found in very massive clusters
(Cassano \& Brunetti \cite{cas05}; Cassano et al. \cite{cas06}).  Very
recently, a unified halo-relic model has been presented in the
framework of hadronic models where the time-dependence of the
magnetic fields and of the cosmic ray distributions is taken into
account to explain the observational properties of both halos and
(most) relics (Keshet \cite{kes10}). In this model the ICM
magnetization is triggered by a merger event, in part but probably not
exclusively in the wake of merger shocks.

Unfortunately, one has been able to study these phenomena only
recently on the basis of a sufficient statistics, i.e.  few dozen
clusters hosting diffuse radio sources up to $z\sim 0.5$ (e.g.,
Giovannini et al. \cite{gio99}; see also Giovannini \& Feretti
\cite{gio02}; Feretti \cite{fer05}; Venturi et al. \cite{ven08};
Bonafede et al. \cite{bon09}; Giovannini et al. \cite{gio09}).  It is expected that new radio telescopes will
largely increase the statistics of diffuse sources (e.g., LOFAR,
Cassano et al. \cite{cas10a}) and allow the study of diffuse radio
emission in low X-ray luminosity clusters to discriminate among
theories of halo formation (e.g., Cassano et al. \cite{cas05}; Ensslin
et al. \cite{ens11}).  The study of galaxy clusters with radio
emission offers a unique tool to estimate strength and structure of
large-scale magnetic fields and might have important cosmological
implications (see Dolag et al. \cite{fer08} and Ferrari et
al. \cite{fer08} for recent reviews). In particular, the study of
clusters with radio halos/relics will likely contribute to quantify
the effect of the non-thermal pressure to the estimate of mass and
temperature in galaxy clusters (e.g., Loeb \& Mao \cite{loe94};
Markevitch \cite{mar10}) and, more in general, the thermal and non
thermal effects of cluster mergers on global properties and
cosmological parameters (e.g., Sarazin \cite{sar04}).

From the observational point of view, there is growing evidence of the
connection between diffuse radio emission and cluster mergers, since
up to now diffuse radio sources have been detected only in merging
systems (see Cassano et al. \cite{cas10b}). In most cases the cluster
dynamical state has been derived from X-ray observations (Schuecker
et al. \cite{sch01}; Buote \cite{buo02}; Cassano et al.
\cite{cas10b}).  Optical data are a powerful way to investigate the
presence and the dynamics of cluster mergers, too (e.g., Girardi \&
Biviano \cite{gir02}). The spatial and kinematical analysis of member
galaxies allow us to reveal and measure the amount of substructure,
and to detect and analyze possible pre-merging clumps or merger
remnants.  This optical information is really complementary to X-ray
information since galaxies and the ICM react on different timescales
during a merger (see, e.g., the numerical simulations by Roettiger et
al. \cite{roe97}).  In this context, we are conducting an intensive
observational and data analysis program to study the internal dynamics
of clusters with diffuse radio emission by using member galaxies (DARC
-- Dynamical Analysis of Radio Clusters -- project, see Girardi et
al. \cite{gir07}\footnote{see also the web site of the DARC project
  http://adlibitum.oat.ts.astro.it/girardi/darc.}).

During our observational program, we have conducted an intensive study
of the cluster \object{Abell 2254} (hereafter A2254).  A2254 is a very
rich, X-ray luminous cluster: Abell richness class $=2$ (Abell et
al. \cite{abe89}); $L_\mathrm{X}$(0.1--2.4 keV)=7.19$\times 10^{44}
\ h_{50}^{-2}$ erg\ s$^{-1}$ (Ebeling et al. \cite{ebe96}). Optically,
the cluster is classified as a Rood-Sastry morphological type ``B'',
i.e. binary (Struble \& Rood \cite{str87}). The cluster redshift
reported in the literature refers to that $z=0.178$ of the brightest
cluster galaxy (Crawford et al.~\cite{cra95}). 

Evidence for the existence of an extended, diffuse radio source was
reported by Owen et al. (\cite{owe99}) and the presence of a radio
halo, having a projected size of $\sim 5$\arcm, is unambiguously shown
by Giovannini et al. (\cite{gio99}).  Both the radio and the X-ray
emission (VLA 1.4 GHz and ROSAT HRI data, respectively) show a very
clumpy and irregular structure (Govoni et al. \cite{gov01b}), with a
radio power at 1.4 GHz of $P_{\rm 1.4GHz}=2.9\times10^{24}\ h_{70}^{-2}$ W\ Hz$^{-1}$ and a large
linear size of $\sim 0.9$ \hh.

We included this cluster in our DARC sample and obtained new
spectroscopic and photometric data from the Telescopio Nazionale
Galileo (TNG) and the Isaac Newton Telescope (INT), respectively. Our
present analysis is based on these optical data, Subaru imaging data
and XMM-Newton Science Archive data, too.

This paper is organized as follows. We present optical data and the
cluster catalog in Sect.~2. We present our results about the cluster
structure based on optical and X-ray data in Sects.~3 and 4,
respectively.  We discuss our results and present our conclusions in
Sect.~5.

\begin{figure*}
\centering 
\includegraphics[width=18cm]{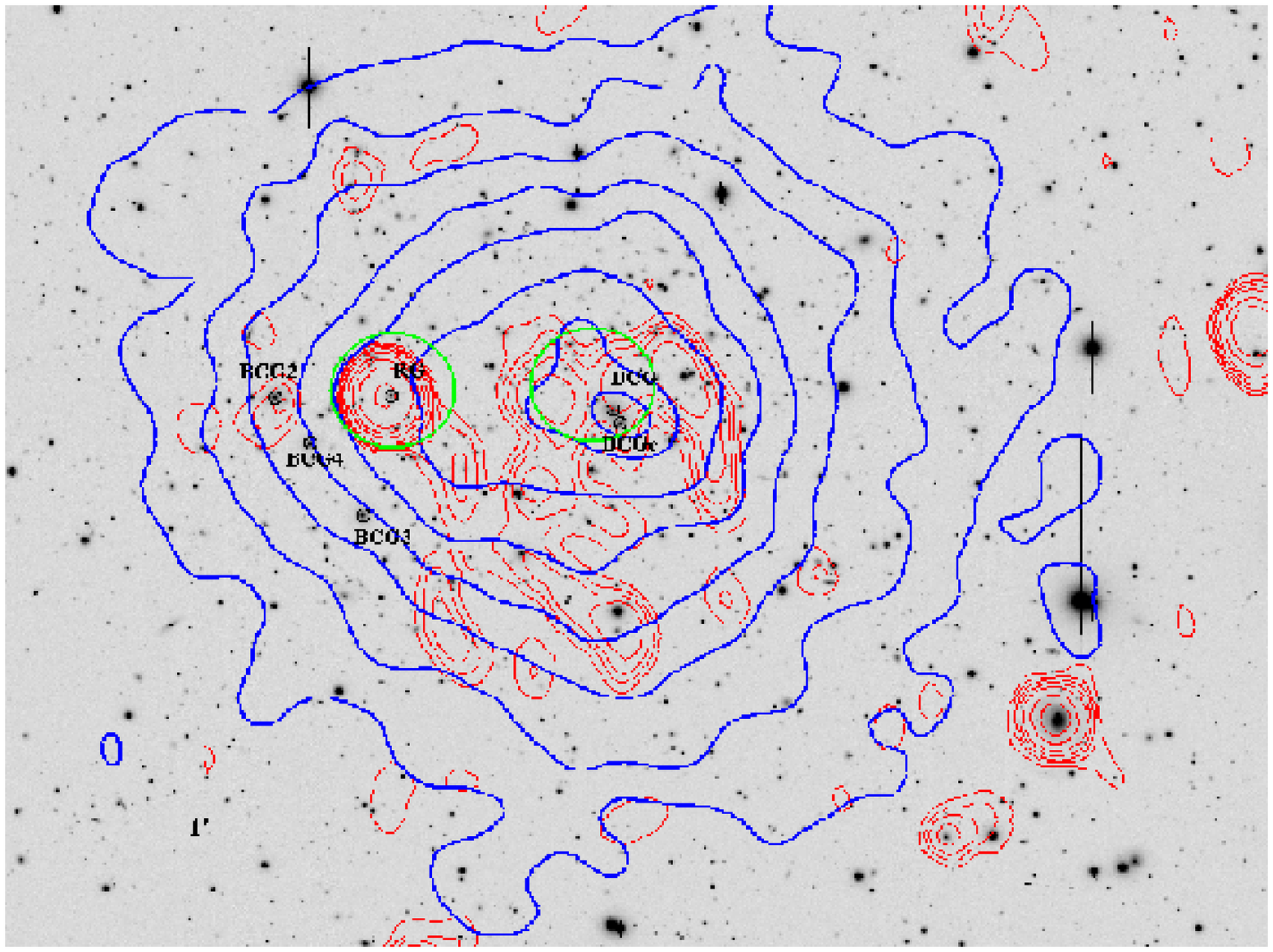}
\caption{Subaru $V$-band image of the cluster A2254 (North at the top
  and East to the left) with, superimposed, the contour levels of the
  XMM archival image (thick/blue contours; photons in the energy range
  0.5--2 keV) and the contour levels of a VLA radio image at 1.4 GHz
  (thin/red contours, see Govoni et al. \cite{gov01b}). Labels
  indicate galaxies cited in the text. The two (green) circles
indicate the positions of the E- and W-peaks in the galaxy distribution
from the 2D-DEDICA analysis (Subaru sample, with $i^{\prime}\le 20.5$ mag,
see Tab.~\ref{tabdedica2d}).
}
\label{figimage}
\end{figure*}
 
Unless otherwise stated, we indicate errors at the 68\% confidence
level (hereafter c.l.).  Throughout this paper, we use $H_0=70$ km
s$^{-1}$ Mpc$^{-1}$ and $h_{70}=H_0/(70$ km s$^{-1}$ Mpc$^{-1}$) in a
flat cosmology with $\Omega_0=0.3$ and $\Omega_{\Lambda}=0.7$. In the
adopted cosmology, 1\arcm corresponds to $\sim 180$ \kpc at the
cluster redshift.

\begin{figure*}
\centering 
\includegraphics[width=18cm]{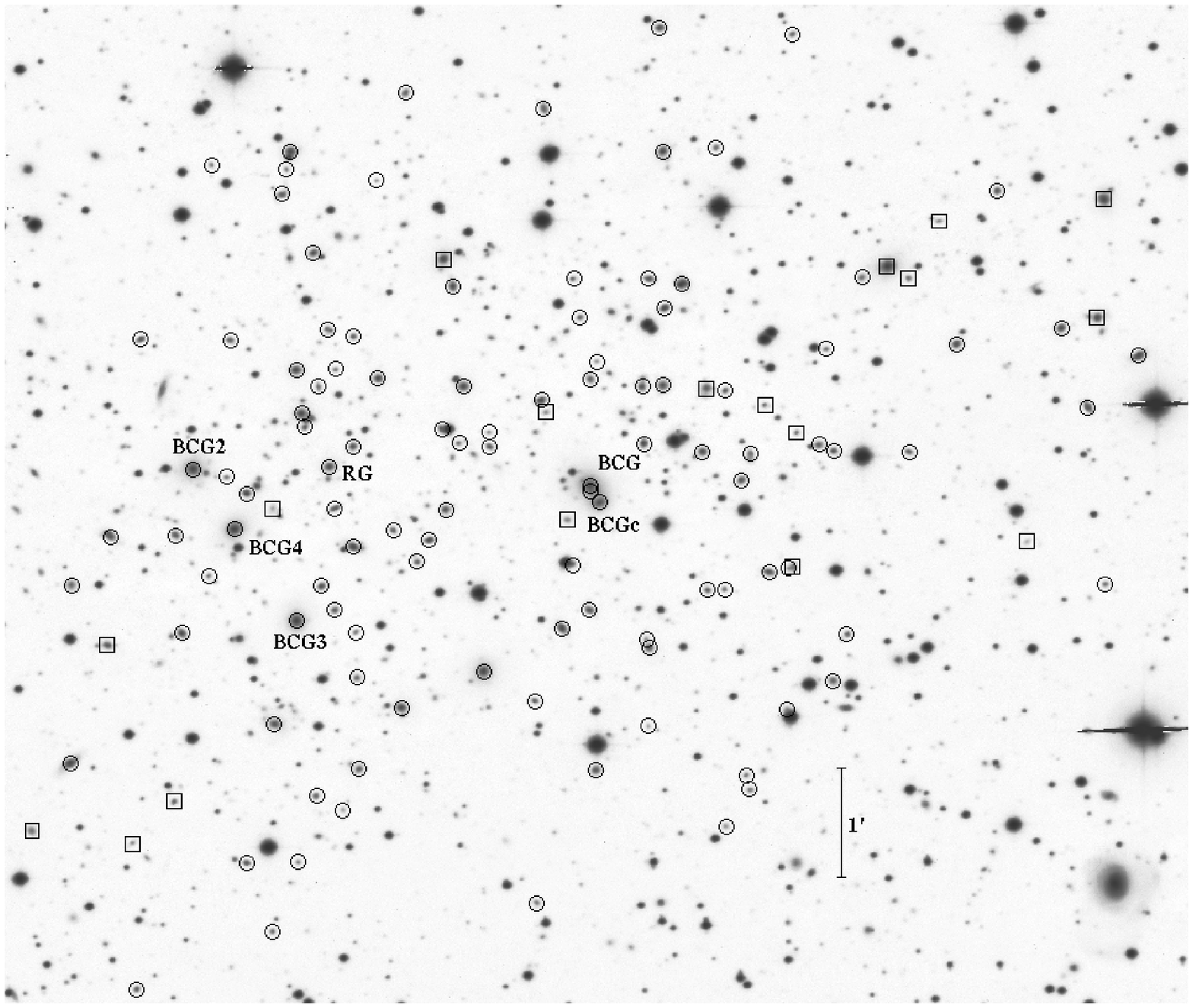}
\caption{INT $r^{\prime}$-band image of the cluster A2254 (North at
  the top and East to the left). Circles and squares indicate cluster
  members and non-members, respectively (see
  Table~\ref{catalogA2254}).  Labels indicate cluster galaxies cited
  in the text.}
\label{figottico}
\end{figure*}

\section{Galaxy data and catalog}
\label{data}

\subsection{Spectroscopic observations}
\label{spec}

Multi-object spectroscopic observations of A2254 were carried out at
the TNG telescope in May/October 2009 and in March 2010. We used
DOLORES/MOS with the LR-B Grism 1, yielding a dispersion of 187
\AA/mm.  We used the $2048\times2048$ pixels E2V CCD, with a pixel
size of 13.5 $\mu$m. In total, we observed 4 MOS masks for a total of
142 slits. Total exposure times were 3600 s for two masks and 5400 s
for the other two masks.

Wavelength calibration was performed using helium and neon-mercury
lamps. Reduction of spectroscopic data was carried out using the
IRAF\footnote{IRAF is distributed by the National Optical Astronomy
  Observatories, which are operated by the Association of Universities
  for Research in Astronomy, Inc., under cooperative agreement with
  the National Science Foundation.} package. Radial velocities were
determined using the cross-correlation technique (Tonry \& Davis
\cite{ton79}) implemented in the RVSAO package (developed at the
Smithsonian Astrophysical Observatory Telescope Data Center). Each
spectrum was correlated against five templates for a variety of galaxy
spectral types: E, S0, Sa, Sb, and Sc (Kennicutt \cite{ken92}). The
template producing the highest value of $\cal R$, i.e., the parameter
given by RVSAO and related to the signal-to-noise ratio of the
correlation peak, was chosen. Moreover, all spectra and their best
correlation functions were examined visually to verify the redshift
determination. In three cases (IDs.~24, 35 and 60; see
Table~\ref{catalogA2254}), we assumed the EMSAO redshift to be a
reliable estimate of the redshift.
Our spectroscopic catalog lists 128 galaxies in the field of A2254.

The formal errors as given by the cross-correlation are known to be
smaller than the true errors (e.g., Malumuth et al. \cite{mal92};
Bardelli et al. \cite{bar94}; Ellingson \& Yee \cite{ell94}; Quintana
et al. \cite{qui00}). Duplicate observations for the same galaxy
allowed us to estimate the true intrinsic errors in data of the same
quality taken with the same instrument (e.g. Barrena et
al. \cite{bar09}).  Here we have double determinations
for seven galaxies, thus we decided to apply the procedure that had
already been applied in the above study obtaining that true
errors are larger than formal cross-correlation errors by a factor
of 2.2. For the seven galaxies with two redshift estimates, we used
the mean of the two measurements and the corresponding
errors. As for the radial velocities estimated through EMSAO we
assumed the largest between the formal error and 100 \kss.  As for
the whole catalog, the median value of the $cz$ errors, taking into account
the above correction, is 66 \kss.

\subsection{INT observations}
\label{int}

Our photometric observations were carried out with the Wide Field
Camera (WFC), mounted at the prime focus of the 2.5m INT telescope. We
observed A2254 in $g^{\prime}$, $r^{\prime}$ and $i^{\prime}$ SDSS
filters in photometric conditions and a seeing of $\sim$1.6\arcss.

The WFC consists of a four-CCD mosaic covering a
33\arcmm$\times$33\arcm field of view, with only a 20\% marginally
vignetted area. We took twelve exposures of 600 s in $g^{\prime}$
filter, nine exposures of 450 s in $r^{\prime}$ filter and nine
exposures of 400 s in $i^{\prime}$ filter. So a total of 7200 s in
$g^{\prime}$ filter, 4050 s in $r^{\prime}$ and 3600 s in $i^{\prime}$
band. The observations were performed making a dithering pattern. This
observing procedure allowed us to build a ``supersky'' frame that was
used to correct our images for fringing patterns (Gullixson
\cite{gul92}). In addition, the dithering helped us to clean cosmic
rays and avoid the effects of gaps between the CCDs in the final full
mosaic images.  Another problem associated with the wide field frames
is the distortion of the field. To match the photometry of several
filters, a good astrometric solution is needed to take into account
these distortions. Using the $imcoords$ IRAF tasks and taking as a
reference the USNO B1.0 catalog, we were able to find an accurate
astrometric solution (rms $\sim$0.3\arcss) across the full frame. Only
in the North-East edge of the image, at $\sim$17\arcm from the
cluster center, we could not obtain a sufficiently good astrometric
solution. In order to guarantee a good match between $g^{\prime}$,
$r^{\prime}$ and $i^{\prime}$ catalogs, this marginal region was
removed from the master catalog.  The photometric calibration was
performed by observing the SA108 standard Landolt field (Landolt
\cite{lan92}) that is also calibrated in the SDSS photometric system
(Smith \cite{smi02}).

We finally identified galaxies in our $g^{\prime}$, $r^{\prime}$ and
$i^{\prime}$ images and measured their magnitudes with the SExtractor
package (Bertin \& Arnouts \cite{ber96}) and AUTOMAG procedure.
Objects were identified by imposing that they cover a certain minimum
area and have a number of counts above a limiting threshold taking the
sky local background as a reference. The limiting size and flux were
16 pixels and 1.5 standard deviation above the sky counts level,
respectively. The selected limiting size corresponds to an apparent
size of 1.3\arcsec, which is about the minimum seeing size during the
observations. We also performed visual inspections of the frames in
order to deal with the best combination of the above parameters that
remove spurious objects from the catalogs.

In a few cases (e.g., close couples of galaxies, galaxies very close to
bright stars or close to defects of the CCD) the standard SExtractor
photometric procedure failed. In these cases, we computed magnitudes
by hand. This method consisted of measuring fluxes assuming a galaxy
profile of a typical elliptical galaxy and scaling it to the maximum
observed value. The integration of this profile provided an estimate
of the magnitude. This method is similar to PSF photometry, but
assumes an E-type galaxy profile, which is more appropriate in this
case.

As a final step, we estimated and corrected the Galactic extinction
$A_{g^{\prime}} \sim0.21$, $A_{r^{\prime}} \sim0.15$ and
$A_{i^{\prime}} \sim0.10$ using Schlegel et al. (\cite{sch98})
reddening and extinction maps computed from IRAS and COBE/DIRBE
data. We estimated that our photometric sample is complete down to
$g^{\prime}=23.4$ (24.3), $r^{\prime}=22.1$ (23.3) and
$i^{\prime}=21.9$ (23.3) for $S/N=5$ (3) within the observed field.
We assigned magnitudes to all galaxies of our spectroscopic catalog.

\subsection{Subaru images}
\label{Subaru}

Subaru Suprime-Cam data were retrieved from the SMOKA
archive\footnote{http://smoka.nao.ac.jp/SUPsearch}. Following the same
steps adopted by Nonino et al. (\cite{non09}), the images in $V$-band
and $i^{\prime}$-band were coadded and reduced to cover a
30\arcmm$\times$30\arcm field of view.  Seeing conditions were $\sim$
0.8\arcsec -- 1.0\arcsec, with a total exposure time of 1200s in
$V$-band, and 1920s in $i^{\prime}$-band. The astrometric solutions
were obtained using 2MASS as reference.

Unfortunately, the observations in both filters were collected under
non-photometric conditions, as shown from the relative photometric
analysis. Therefore, the final coadded images were not calibrated
using standard stars. For the use we make in this study, i.e. the 2D
analysis of substructure, non calibrated magnitudes are quite
sufficient and we refer them as $i^{\prime}_{\rm obs}$ and $V_{\rm
  obs}$ (see Sect.~\ref{photo} for other details).

\subsection{Galaxy catalog}
\label{cat}

Table~\ref{catalogA2254} lists the velocity catalog (see also
Fig.~\ref{figottico}): identification number of each galaxy and member
galaxies, ID and IDm (Cols.~1 and 2, respectively); right ascension
and declination, $\alpha$ and $\delta$ (J2000, Col.~3); $g^{\prime}$,
$r^{\prime}$, and $i^{\prime}$ INT magnitudes (Cols.~4, 5 and 6);
heliocentric radial velocities, ${\rm v}=cz_{\sun}$ (Col.~7) with
errors, $\Delta {\rm v}$ (Col.~8).

\input{catalogA2254.tex}

\subsection{Individual galaxies}
\label{gals}

No evident dominant galaxy is present in the cluster since the
brightest galaxy in our catalog (ID.~53, $r^{\prime}=16.22$, hereafter
BCG) is less than 0.5 mag brighter then the following brightest
galaxies (IDs.~117, 102, 112 with $r^{\prime}=16.64$, 16.66, and 16.67
respectively).  These bright galaxies all lie in the eastern region of
the cluster; we refer to them as BCG2, BCG3, and BCG4.

The BCG is close to the peak of the X-ray emission and has a
(projected) bright companion ($r^{\prime}=17.01$, ID.~50 hereafter
BCGc).  BCGc is characterized by [OII], H$\beta$, and [OIII] emission
spectral lines, strong $24\mu$ MIPS-Spitzer emission\footnote{see at
  http://sha.ipac.caltech.edu/applications/Spitzer/SHA/}, UV
emission\footnote{http://galex.stsci.edu/GalexView/}, and faint radio
emission (ID.~2 in Table~3 of Rizza et al. \cite{riz03}). The Subaru
images show that it is a spiral galaxy with a large bulge and an arm
directed toward the BCG (see Fig.~\ref{merging}). We find a quite
large relative velocity difference with respect to the BCG ($\sim
1400$ \kss).

Figure~2 of Govoni et al. (\cite{gov01b}) suggests the presence of a
few discrete radio sources. The brightest one -- with a flux of $\sim$
38 mJy at 1.4 GHz -- lies in the NE cluster region where the field is
very crowded. The likely optical counterpart might be the cluster
galaxy ID.~93 (hereafter ``RG'') which has an optical spectrum typical
of ellipticals.  ID.~93, with an absolute magnitude $M_{R}\sim -21.9$
and a radio power log $P_{\rm 1.4GHz}({\rm W\,Hz}^{-1})\sim 24.5$, can
be accommodated among the radio galaxies in clusters (e.g., Ledlow \&
Owen \cite{led96}; Bardelli et al. \cite{bar10}).  However, a very
faint, extended object close to RG is revealed by the Subaru image
(Fig.~\ref{agn}). Due to the present resolution in radio data, there
is some doubt on the identification of RG as the (unique) responsible
of the whole radio emission of the bright discrete radio source.
Figure~\ref{agn} also shows the position of a point-like X-ray
source and of a MIPS-Spitzer source.

A full discussion of individual galaxies is far from the aim of this
study and we only give Figs.~\ref{merging} and ~\ref{agn} in the
electronic version of the paper.

\onlfig{3}{
\begin{figure}
\centering
\resizebox{\hsize}{!}{\includegraphics{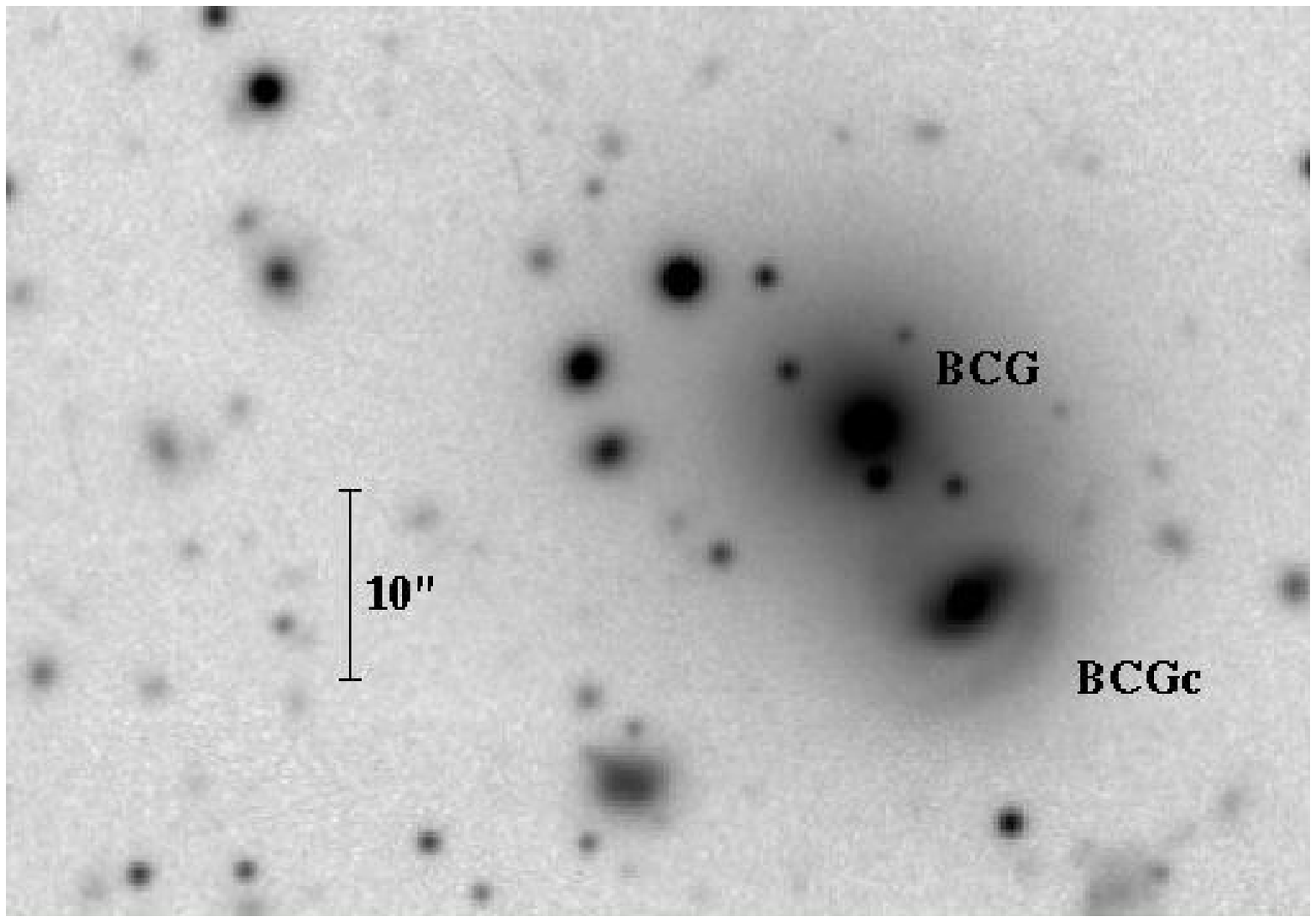}}
\caption{Subaru $i^{\prime}$-band image of the very central cluster
  region. It highlights the BCG and its bright spiral companion BCGc
  (North at the top and East to the left).}
\label{merging}
\end{figure}
}

\onlfig{4}{
\begin{figure}
\centering
\resizebox{\hsize}{!}{\includegraphics{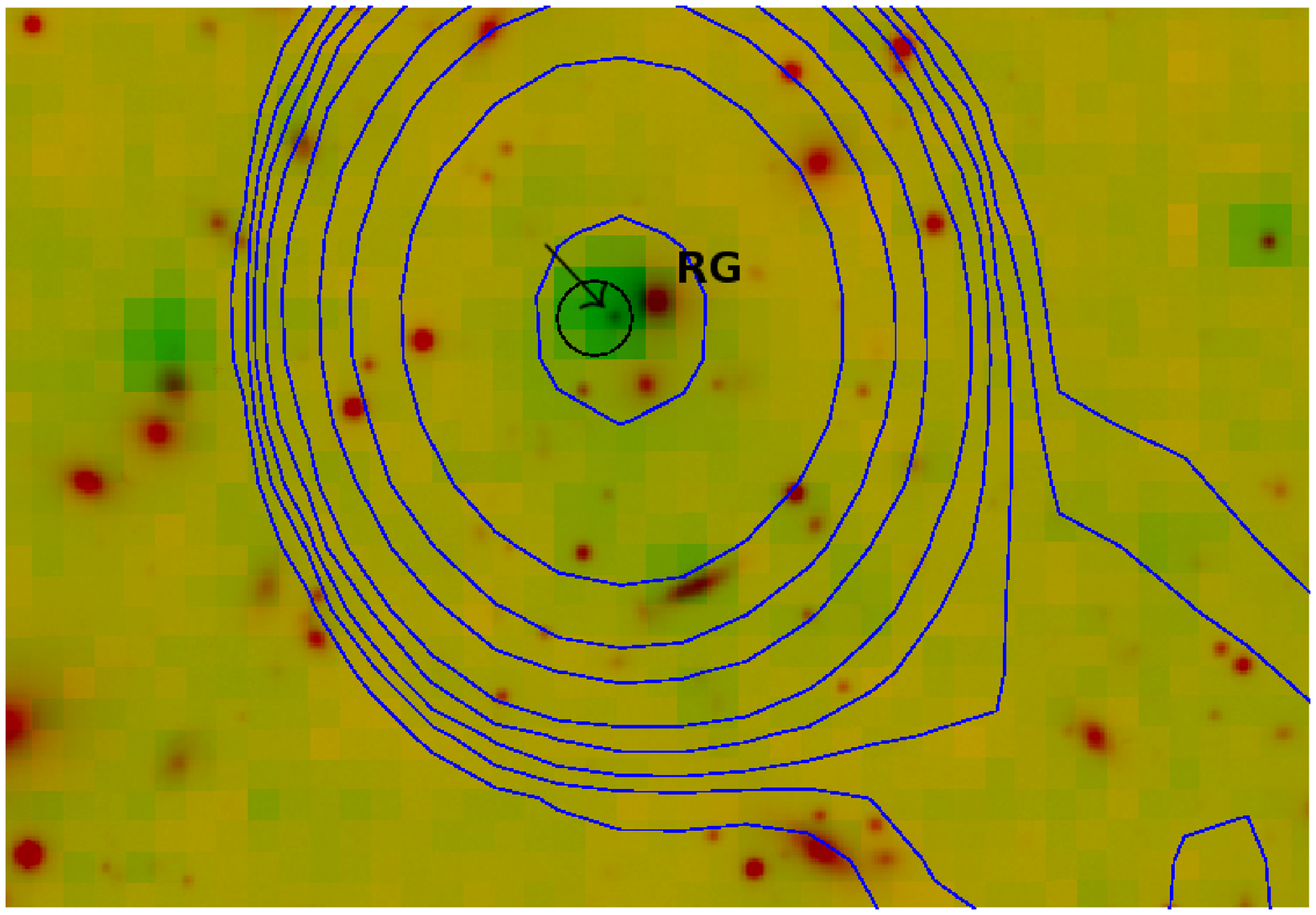}}
\caption{Multiwavelength picture of the region around RG galaxy.  The
  MIPS-Spitzer image (green color) is superimposed to the Subaru
  $i^{\prime}$-band image. The arrow indicates the faint extended
  object close to RG.  The black circle indicate the position of the
  XMM point source. Contour levels of the radio image are shown in
  blue (Govoni et al. \cite{gov01b}). North at the top and East to the
  left.}
\label{agn}
\end{figure}
}

\section{Analysis of the optical data}
\label{anal}

\subsection{Member selection}
\label{memb}

\begin{figure}
\centering
\resizebox{\hsize}{!}{\includegraphics{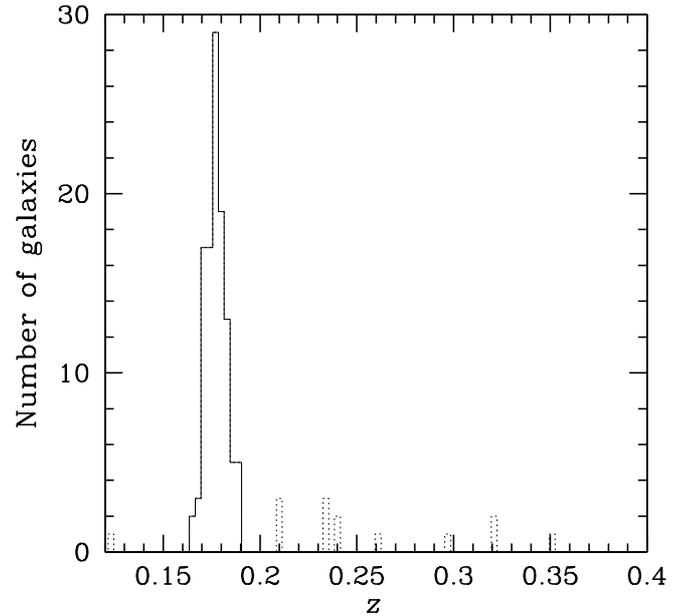}}
\caption
{Redshift galaxy distribution. The solid line histogram refers to the
  110 galaxies assigned to A2254 according to the DEDICA
  reconstruction method.}
\label{fighisto}
\end{figure}

To select cluster members among the 128 galaxies with redshifts we
perform the 1D adaptive-kernel method (hereafter DEDICA, Pisani
\cite{pis93} and \cite{pis96}; see also Fadda et al. \cite{fad96};
Girardi et al. \cite{gir96}). We search for significant peaks in the
velocity distribution at $>$99\% c.l.. This procedure detects A2254 as
a peak at $z\sim0.177$ populated by 110 galaxies considered as
fiducial cluster members (in the range $49\,237\leq {\rm v} \leq
56\,757$ \kss, see Fig.~\ref{fighisto}).  As for the center of A2254,
we adopt the position of the BCG
[R.A.=$17^{\mathrm{h}}17^{\mathrm{m}}45\dotsec86$, Dec.=$+19\degree
  40\arcmm 48.4\arcs$ (J2000.0)].  The 18 non-members are 2 and 16
foreground and background galaxies, respectively.

\begin{figure}
\centering
\resizebox{\hsize}{!}{\includegraphics{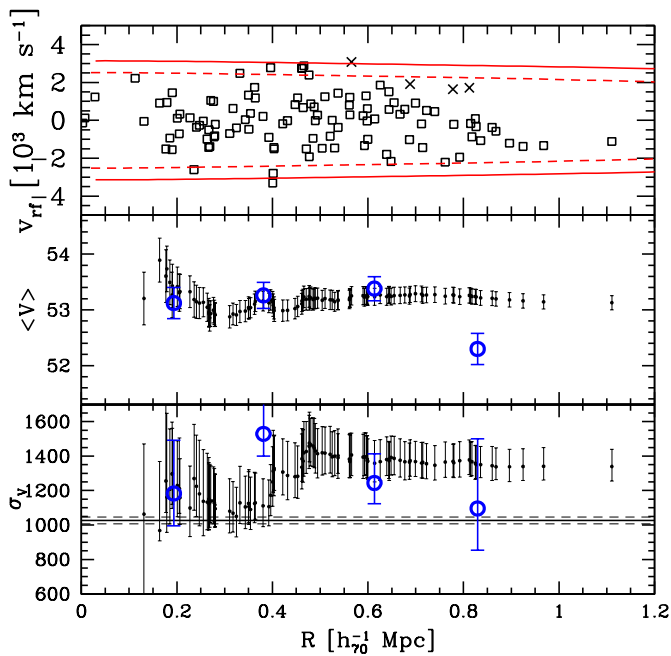}}
\caption
{{\em Top panel:} rest-frame velocity vs. projected clustercentric
  distance for the 110 member galaxies of the fiducial sample
  (Fig.~\ref{fighisto}). Crosses indicate galaxies formally rejected
  as interlopers by the ``shifting gapper'' procedure, but here
  considered in a more conservative view.  The cluster center
  coincides with the position of the BCG.  Red solid (dashed)
    curves enclose the region of allowed values for member galaxies
    according to the criterion of den Hartog \& Katgert (\cite{den96})
    and using the upper (lower) limit of our mass estimate computed in
    the following (see Sect.~\ref{mass}).  {\em Middle and bottom
    panels:} differential (big circles) and integral (small points)
  profiles of mean velocity and LOS velocity dispersion, respectively.
  For the differential profiles, we plot the values for four annuli
  from the center of the cluster, each of 0.25 \h (large blue
  symbols).  For the integral profiles, the mean and dispersion at a
  given (projected) radius from the cluster-center is estimated by
  considering all galaxies within that radius -- the first value
  computed on the five galaxies closest to the center. The error bands
  at the $68\%$ c.l. are also shown.  In the bottom panel, the
  horizontal line represents the X-ray temperature with the respective
  errors transformed in $\sigma_{\rm V}$ assuming the density-energy
  equipartition between ICM and galaxies, i.e.  $\beta_{\rm spec}=1$
  (see Sect.~\ref{disc}).  }
\label{figprof}
\end{figure}

In the previous analyses of DARC clusters we generally used a second
step in the member selection, which uses the combination of position
and velocity information, i.e., the ``shifting gapper'' method by
Fadda et al. (\cite{fad96}). This procedure rejects galaxies that
  are too far in velocity from the main body of galaxies within a
  fixed bin that shifts along the distance from the cluster center
  (adopted values are $1000$ \ks in the cluster rest-frame for the
  velocity cut and 0.6 \hh, or large enough to include 15 galaxies,
  for the spatial shifting bin).  The procedure is iterated until the
  number of cluster members converges to a stable value (see Fadda et
  al. for other details and their Fig.~1 for an example).  Here, the
formal application of the ``shifting gapper'' procedure would reject
another four galaxies just at the border of the distribution of
galaxies in the plane of the rest-frame velocity vs. projected
clustercentric distance (see the crosses in the Fig.~\ref{figprof} --
top panel). These galaxies are not spatially clustered and have the
typical early-type galaxy colors. Therefore, we decide to not reject
them from the analysis.  Global cluster quantities are independent of
this decision within the 1$\sigma$ error.  Note that the
  ``shifting gapper'' procedure is completely empirical and has the
  advantage of being independent of hypotheses about the poorly known
  dynamical status of the cluster. Other available procedures
  combining position and velocity information are generally based on
  physical assumptions about the dynamical status of the cluster and
  the consequent possibility to apply the virial mass and/or the
  knowledge of the mass profile (e.g., den Hartog \& Katgert
  \cite{den96}, see Biviano et al. \cite{biv06} for a recent
  application). Therefore, the application of these procedures is not
  well justified in the case of A2254 showing evidence of substructure
  (see in the following). Making an attempt in this direction, we use
  the value of the mass obtained in Sect.~\ref{mass}, $M_{\rm sys}$,
  and adopt a one-single cluster model described by the NFW mass
  profile (Navarro et al. \cite{nav97}, see Sect.~\ref{mass} too) to
  construct the upper limits to the LOS velocities at projected
  distance $R$ from the cluster center (criterion (i) by den Hartog \&
  Katgert \cite{den96}).  Figure~\ref{figprof} shows as the region of
  allowed velocities is roughly coincident with that occupied by our
  selected member galaxies. Thus, although in the case of the
  substructured cluster A2254 we do not expect to determine the
  detailed cluster membership, it supports the consistency between our
  adopted galaxy membership and the value we obtain for the mass
  estimate.

\subsection{Global cluster properties}
\label{glob}

By applying the biweight estimator to the 110 cluster members (Beers et
al. \cite{bee90}, ROSTAT software), we compute a mean cluster redshift
of $\left<z\right>=0.1772\pm$ 0.0004, i.e.
$\left<\rm{v}\right>=(53\,128\pm$125) \kss.  We estimate the LOS
velocity dispersion, $\sigma_{\rm V}$, by using the biweight estimator
and applying the cosmological correction and the standard correction
for velocity errors (Danese et al. \cite{dan80}).  We obtain
$\sigma_{\rm V}=1340_{-84}^{+101}$ \kss, where errors are estimated
through a bootstrap technique.

To evaluate the robustness of the $\sigma_{\rm V}$ estimate, we
analyze the velocity dispersion profile (Fig.~\ref{figprof}). The
integral profile shows an enhancement at $\sim0.4$--0.5 \h from the
cluster center and then flattens. This suggests that a robust value of
$\sigma_{\rm V}$ is asymptotically reached in the external cluster
regions, as found for most nearby clusters (e.g., Fadda et
al. \cite{fad96}; Girardi et al. \cite{gir96}).  As for the
enhancement at $\sim0.4$--0.5 \hh, it is likely due to the
contamination of the galaxies of a secondary clump (see
Sects.~\ref{photo} and \ref{3d}); see also, e.g., Girardi et
al. (\cite{gir96}) and Barrena al. (\cite{bar07}) for examples in
other clusters).

\subsection{Velocity distribution}
\label{velo}

We analyze the velocity distribution to search for possible deviations
from Gaussianity that might provide important signatures of complex
dynamics. For the following tests, the null hypothesis is that the
velocity distribution is a single Gaussian.

We estimate three shape estimators, i.e., the kurtosis, the skewness,
and the scaled tail index (see, e.g., Bird \& Beers \cite{bir93}).
There is no evidence of departures from the Gaussianity (see Table~2
of Bird \& Beers \cite{bir93}).

\begin{figure}
\centering 
\resizebox{\hsize}{!}{\includegraphics{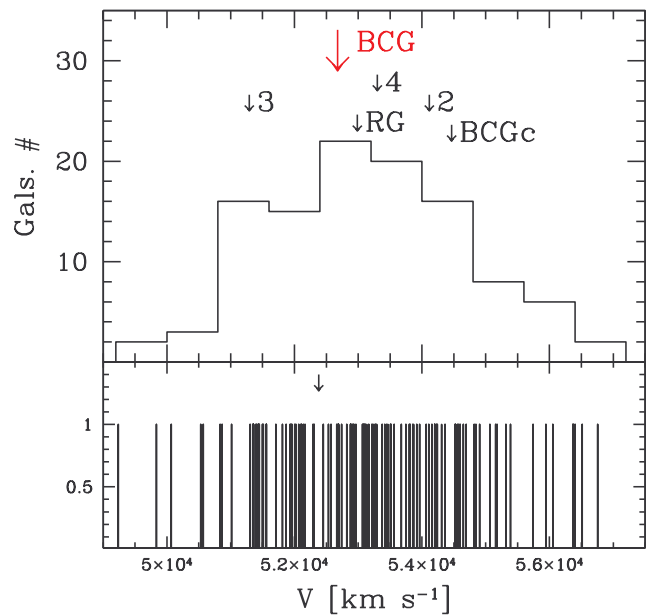}}
\caption
{The 110 galaxies assigned to the cluster.  {\em Upper panel}:
  Velocity distribution.  The arrows indicate the velocities of the
  BCG, other brightest galaxies, and RG.  {\em
    Lower panel}: Stripe density plot where the arrow indicates the
  position of the significant gap.}
\label{figstrip}
\end{figure}

We then investigate the presence of gaps in the velocity distribution.
We follow the weighted gap analysis presented by Beers et
al. (\cite{bee91}; \cite{bee92}; ROSTAT software).  We look for
normalized gaps larger than 2.25 since in random draws of a Gaussian
distribution they arise at most in about $3\%$ of the cases,
independent of the sample size (Wainer and Schacht~\cite{wai78}). We
detect one significant gap (at the $97\%$ c.l.), which divide the
cluster into two groups of 36 and 70 galaxies from low to high
velocities (hereafter GV1 and GV2, see Fig.~\ref{figstrip}).  The BCG3
is assigned to the GV1 peak.  Other bright galaxies (BCG, BCG2, BCG4,
and BCGc) are all assigned to the GV2 peak.

To detect subsets in the velocity distribution we then resort to the
Kaye's mixture model (KMM) test (Ashman et al. \cite{ash94}).  The KMM
algorithm fits a user-specified number of Gaussian distributions to a
dataset and assesses the improvement of that fit over a single
Gaussian. In addition, it provides the maximum-likelihood estimate of
the unknown n-mode Gaussians and an assignment of objects into
groups. The KMM test is more appropriate in situations where
theoretical and/or empirical arguments indicate that a Gaussian model
is reasonable.  The Gaussian is valid in the case of cluster velocity
distributions, where gravitational interactions drive the system
toward a relaxed configuration with a Gaussian velocity
distribution. The KMM test does not find a two-group partition, which
provides a significantly more accurate description of the velocity
distribution than a single Gaussian.

To search for a possible physical meaning of the two subclusters
determined by the detected significant velocity gap, we also compare
two by two the spatial galaxy distributions of GV1 and GV2.  We find
no difference according to the the 2D Kolmogorov-Smirnov test (Fasano
\cite{fas87}).

\subsection{Analysis of the 2D galaxy distribution}
\label{photo}

\begin{figure}
\includegraphics[width=8cm]{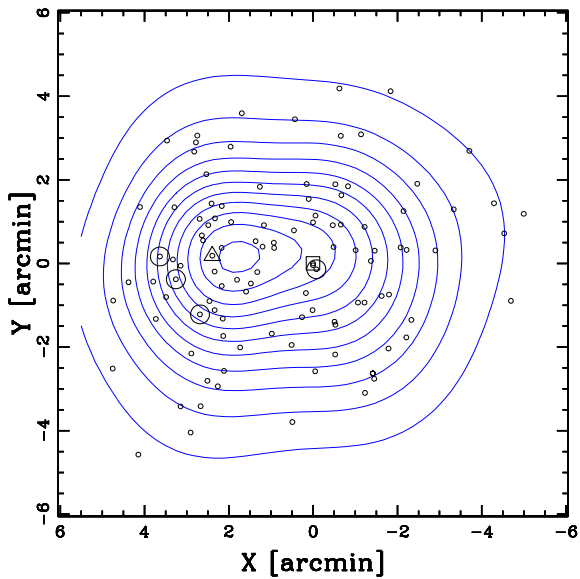}
\caption
{Spatial distribution on the sky and relative isodensity contour map
  of spectroscopic cluster members, obtained with the 2D-DEDICA
  method.  The BCG is taken as the cluster center.  Large square
and circles indicate the location of the BCG and other
brightest galaxies. Large triangle indicates RG.
}
\label{figk2z}
\end{figure}

By applying the 2D adaptive-kernel method (2D-DEDICA) to the
positions of A2254 galaxy members, we find one peak which is $\sim
2$\arcm far East from the position of the BCG and the isodensity
contours design a E-W elongated structure.  (Fig.~\ref{figk2z}). 

Our spectroscopic data do not cover the entire cluster field and are
affected by magnitude incompleteness.  To overcome these problems, we
use our photometric data samples which cover a larger spatial
region. The INT sample has the advantage to have photometry available
in three magnitude bands, while the Subaru sample allows us to extend
our analysis to fainter galaxies.

\begin{figure}
\centering
\includegraphics[width=8cm]{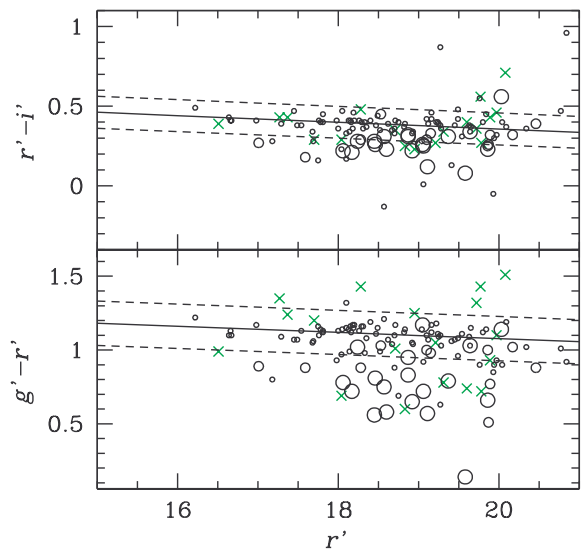}
\caption
{{\em Upper panel:}  INT $r^{\prime}-i^{\prime}$ vs. $r^{\prime}$ diagram
  for galaxies with available spectroscopy.  Black circles and green
  crosses indicate member and no-member galaxies. For member galaxies
  large/middle/small size circles indicate late/middle/early type
  galaxies.  The solid line gives the CMR determined on member
  galaxies; the dashed lines are drawn at $\pm$0.1 mag from this
  value.  {\em Lower panel:}  INT $g^{\prime}-r^{\prime}$ vs. $r^{\prime}$
  diagram for galaxies with available spectroscopy.  The dashed lines
  are drawn at $\pm$0.15 mag from the CMR determined on member
  galaxies.  }
\label{figcm}
\end{figure}

When more than two colors are available, it is more effective to
select galaxies in the color-color space (Goto et al.  \cite{got02}).
In particular, Lu et al. (\cite{lu09}, see their Figs.~6 and 7) show
as ($r^{\prime}$--$i^{\prime}$) color can effectively eliminate high
redshift galaxies from the red-sequence galaxies at $z$=0.2--0.3
redshift.  In our INT photometric catalog we select likely members on
the basis of both ($r^{\prime}$--$i^{\prime}$ vs. $r^{\prime}$) and
($g^{\prime}$--$r^{\prime}$ vs. $r^{\prime}$) color-magnitude
relations (hereafter CMRs), which indicate the early-type galaxies
locus.  To determine the CMR we apply the 2$\sigma$-clipping fitting
procedure to the cluster members: we obtain
$r^{\prime}$--$i^{\prime}$=0.777-0.021$\times$ $r^{\prime}$ and
$g^{\prime}$--$r^{\prime}$=1.497-0.021$\times$ $r^{\prime}$ (see
Fig.~\ref{figcm}).  Out of our photometric catalog we consider as
likely ``red'' cluster members those objects lying within 0.1 and 0.15
mag of the CMRs ($r^{\prime}$--$i^{\prime}$ vs. $r^{\prime}$) and
($g^{\prime}$--$r^{\prime}$ vs. $r^{\prime}$), respectively.   The
  selected magnitude intervals are chosen with different widths to
  follow the different amplitudes of the scatter in the two relations:
  0.1 and 0.15 mag are round values roughly corresponding to $\sim
1.3$ times the error associated to the fitted intercepts.  Note
  that the difference in the scatter between the two above CMRs
is quite expected: for instance,  using SDSS data, Goto et
  al. (\cite{got02}) computed a scatter of 0.040 and 0.081
  for the two above relations, respectively.

As a further check, we associate to each galaxy the galaxy type
producing the highest value of $\cal R$ in the RVSAO procedure and
divide galaxies in early-type (E, S0), middle-type (Sa, Sb), and
late-type (Sc, Irr). The galaxies with $z$ determined via EMSAO are
assigned to late-type galaxies. Although these spectral-types do not
pretend to be an alternative to a complete spectral analysis, they are
useful to show how the selection through the CMR, in particular here
the ($g^{\prime}$--$r^{\prime}$ vs. $r^{\prime}$), is useful to
distinguish between early and late-type galaxies (see Fig.~\ref{figcm}
-- lower panel). Figure~\ref{figcm} shows as the selected
  magnitude intervals seem adequate to select early-type galaxies,
  thus to only use good tracers of the cluster substructure (e.g.,
  Lubin et al. \cite{lub00}) and, above all, to avoid non-member
  galaxies which might bias our 2D analysis.

\begin{figure}
\includegraphics[width=8cm]{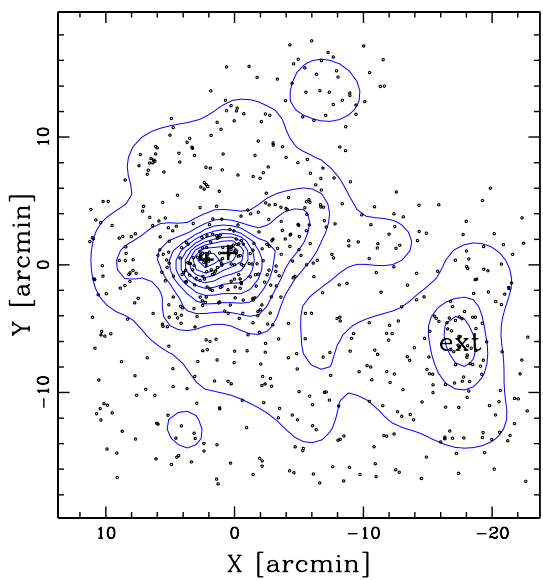}
\caption
{Spatial distribution on the sky and relative isodensity contour map
  of INT photometric cluster members with $r^{\prime}\le 21$, obtained
  with the 2D-DEDICA method. The two crosses indicate the E- and
  W-peaks.}
\label{figk2}
\end{figure}

Figure~\ref{figk2} shows the contour map for the 682 likely cluster
members having $r^{\prime}\le 21$.  We find two, very significant
peaks along the E-W direction, hereafter E- and W-peaks.  The W-peak
is at $\sim 1$\arcm north from the BCG.  A third peak lies in the
WSW region, where our photometric data allow to study very
external cluster regions (hereafter ext-peak).  Similar results are
found for the 780 objects with $r^{\prime}\le 21.5$.

As for the results with $r^{\prime}\le 21$, Table~\ref{tabdedica2d}
lists the number of assigned members, $N_{\rm S}$ (Col.~2); the peak
position (Col.~3); the density (relative to the densest peak),
$\rho_{\rm S}$ (Col.~4); the value of $\chi^2$ for each peak, $\chi
^2_{\rm S}$ (Col.~5).  Ramella et al. (\cite{ram07}) tested the
2D-DEDICA procedure on Monte Carlo simulations reproducing galaxy
clusters. They show that the physical significance, i.e. the
significance which takes into account the noise fluctuations,
associated to the subclusters depends on the statistical significance
of the subcluster (recovered from the $\chi ^2$ value) and can be
computed using simulations.  Considering their eq.~5, the $\chi^2$
threshold for a sample of 682 objects is $\chi^2_{\rm threshold}=$32.
Thus, all the three peaks we detect are physically significant.

\input{tabdedica2d.tex}

\begin{figure}
\centering
\includegraphics[width=8cm]{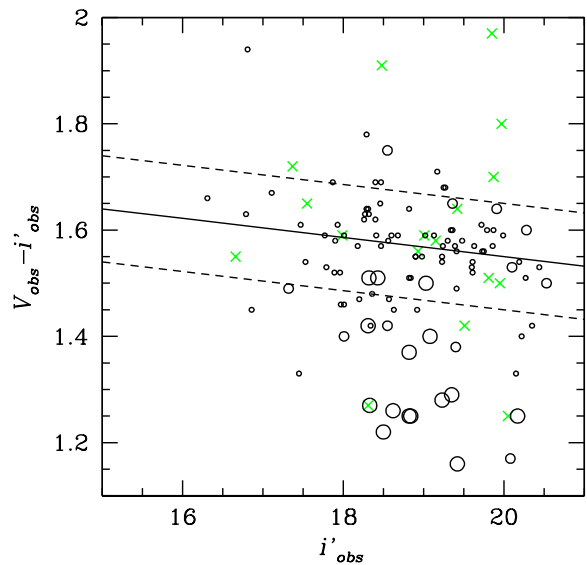}
\caption
{Subaru $V_{obs}-i'_{obs}$ vs. $i'_{obs}$ diagram for galaxies
    with available spectroscopy.  Black circles and green crosses
    indicate member and no-member galaxies. For member galaxies
    large/middle/small size indicate late/middle/early type galaxies.
    The solid line gives the CMR determined on member galaxies; the
    dashed lines are drawn at $\pm$0.1 mag from this value.}
\label{figcmsup}
\end{figure}

In the Subaru photometric catalog we select likely members on the
basis of the ($V_{\rm obs}$--$i^{\prime}_{\rm obs}$)
vs. $i^{\prime}_{\rm obs}$ diagram.  To determine the CMR we apply the
2$\sigma$--clipping fitting procedure to the cluster members: we
obtain $V_{\rm obs}$--$i^{\prime}_{\rm obs}$=1.910--0.018$\times
i^{\prime}_{\rm obs}$.  Out of our photometric catalog, we consider as
likely cluster members the objects lying within 0.1 mag of the CMR
(see Fig.~\ref{figcmsup}). We stress that the use of the above
  color-magnitude relation, fully based on our uncalibrated
  magnitudes, is valid and only valid within our own catalog. However,
  we need a (rough) idea of the magnitude correction to understand the
  deepness of our Subaru samples.  To this aim, we compare our
  uncalibrated magnitudes $i^{\prime}_{\rm obs}$ and magnitudes
  $i^{\prime}$ of the common objects between Subaru catalog and the
  final INT catalog. The visual inspection of the $i^{\prime}_{\rm
    obs}-0.1$ vs. $i^{\prime}$ plot, where 0.1 is the Galactic
  extinction correction, suggests us that $i^{\prime}_{\rm obs}\sim
  i^{\prime}+0.6$ mag.

\begin{figure}
\includegraphics[width=8cm]{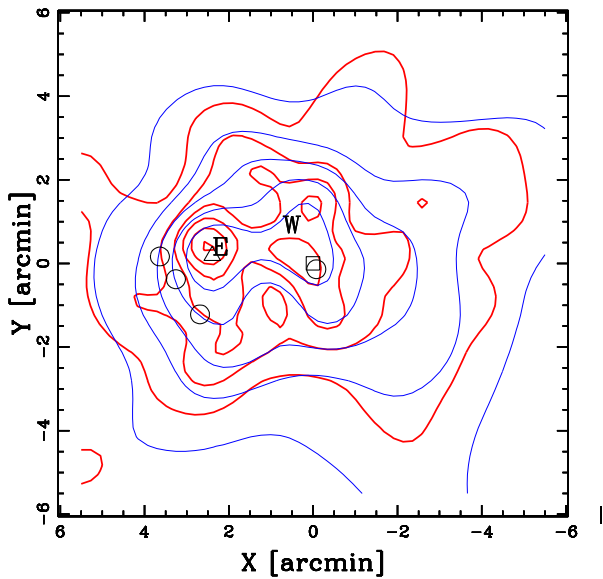}
\caption
{Spatial distribution on the sky and relative isodensity contour map
  of Subaru photometric cluster members with $i^{\prime}\le 20.5$ and
  with $i^{\prime}\le 21.5$ (thin/blue and thick/red contours,
  respectively).  Labels E and W indicate the peaks as detected by
  DEDICA in the INT sample (see Table~\ref{tabdedica2d}).  The BCG and
  other interesting galaxies are indicated as in Fig.~\ref{figk2z}.}
\label{figk2zoom}
\end{figure}

Figure~\ref{figk2zoom} shows the 2D-DEDICA results for the sample
with $i^{\prime}\leq 20.5$, comparable, for magnitude deepness, to the
results from the INT sample with $r^{\prime}\leq 21$.  The
$i^{\prime}\leq 20.5$ sample confirms the presence of two density
peaks in the galaxy distribution of A2254 and the presence of the
external group (841 galaxies in the whole field).  However, while the
E-peak position is close to the INT E-peak, the W-peak is closer to
the BCG position than the INT W-peak and both eastern and western
structures are someway elongated in the N-S direction. These hints
for a more complex structure are reinforced by the analysis of the
$i^{\prime}\leq 21.5$ Subaru sample (Fig.~\ref{figk2zoom}; 1257
galaxies in the whole field) where the W-structure is shown to be
bimodal. Out of many peaks obtained through the 2D-DEDICA in the
$i^{\prime}\leq 21.5$ Subaru sample, Table~\ref{tabdedica2d} lists the
results for the three densest peaks and the external peak for
comparison. Since in this case $\chi^2_{\rm threshold}\sim 40$, we
cannot formally discard the possibility of noise fluctuations, but
Fig.~2 of Ramella et al.  (\cite{ram07}) suggests a small probability
($\sim 10\%$) of noise contamination when the $\chi^2$ is about half
of $\chi^2_{\rm threshold}$.
We do not show results for deeper Subaru samples since, working on
galaxies in magnitudes bins, one realizes that the distribution of
fainter galaxies is no longer centered around the cluster position, a
likely sign of contamination by the background large scale structure.

\begin{figure}
\includegraphics[width=8cm]{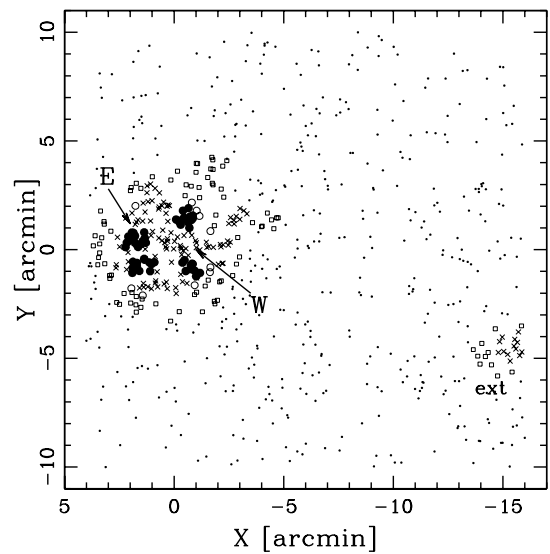}
\caption {Galaxies belonging to structures as detected by the Voronoi
Tessellation and Percolation technique. The algorithm is run on the
sample of likely members with $i^{\prime}<21.5$ extracted from the
Subaru photometric catalog (see text). Open squares, open circles,
crosses and solid circles indicate galaxies in structures at the
95\%, 98\%, 99\% and 99.9\% c. ls., respectively.}
\label{figVTP}
\end{figure}

The complex structure detected by the 2D--DEDICA method is confirmed
by the Voronoi Tessellation and Percolation (VTP) technique
(e.g. Ramella et al. \cite{ram01}; Barrena et al. \cite{bar05}). This
technique is non-parametric and does not smooth the data. As a
consequence, it identifies galaxy structures irrespective of their
shapes. For our purposes we run VTP on the same sample of 1257 Subaru
likely members with $i^{\prime}<21.5$. The result of the
application of VTP is shown in Fig.~\ref{figVTP}. VTP is run four
times adopting four detection thresholds: galaxies identified as
belonging to structures at 95\%, 98\%, 99\% and 99.9\% c.ls. are shown
as open squares, open circles, crosses and solid circles,
respectively. VTP confirms the elongation of the eastern and western
central structures in the N-S direction. In particular, the W-structure is 
clearly bimodal, and there are clues for a bimodality of the E-structure, too.  
Far from the cluster center, $~17$\arcm WSW, VTP also confirms the existence of 
the ext-peak.

Interestingly, the ext--peak can be successfully identified with the
cluster NSC J171627+193456 at z$_{\rm phot}\sim$0.194, discovered by
Gal et al. (\cite{gal03}, based on the galaxy catalogs from the
digitized Second Palomar Observatory Sky Survey).  However, the
absence of spectroscopic $z$ in the region of this system does not
allow to infer if it is a bound companion cluster in the outskirts of
A2254 (projected at $\sim$ 3 $h_{70}^{-1}$ Mpc from the cluster
center).

\subsection{3D analysis}
\label{3d}

\subsubsection{Significance of 3D substructure}
\label{global3d}

The existence of correlations between positions and velocities of
cluster galaxies is a characteristic of true substructures.  Here we
use different approaches to analyze the structure of A2254
combining position and velocity information.

We find no evidence for a significant velocity gradient (see, e.g.,
den Hartog \& Katgert \cite{den96} and Girardi et al. \cite{gir96}
for the details of the method).

In order to check for the presence of substructure, we combine
velocity and position information by computing the $\Delta$-statistics
devised by Dressler \& Schectman (\cite{dre88}, hereafter DS-test),
which is recommended by Pinkney et al.  (\cite{pin96}) as the most
sensitive 3D test.  For each galaxy, the deviation $\delta$ is defined
as $\delta_i^2 =
  [(N_{\rm{nn}}+1)/\sigma_{\rm{V}}^2][(\overline {\rm V_l} - \overline
    {\rm V})^2+(\sigma_{\rm V,l} - \sigma_{\rm V})^2]$, where the
subscript ``l'' denotes the local quantities computed over the
$N_{\rm{nn}}=10$ neighbors of the galaxy.  $\Delta$ is the sum of the
$\delta$ of the individual $N$ galaxies and gives the cumulative
deviation of the local kinematical parameters (mean velocity and
velocity dispersion) from the global cluster parameters.  The
significance of $\Delta$, i.e. of substructure, is checked by running
1000 Monte Carlo simulations, randomly shuffling the galaxy
velocities.  We find no significant presence of substructure.

Following Pinkney et al. (\cite{pin96}; see also Ferrari et
al. \cite{fer03}), we apply two more classical 3D tests: the
$\epsilon$-test (Bird, \cite{bir93}) based on the projected mass
estimator and the centroid shift or $\alpha$-test (West \& Bothun
\cite{wes90}). The details of these tests can be found in the above
papers. We only point out that we consider ten as the number of the
nearest neighbors for each galaxy and we use the above Monte Carlo
simulations to compute the substructure significance.  In both cases
we do not find evidence for significant substructure.

\begin{figure}
\centering 
\resizebox{\hsize}{!}{\includegraphics{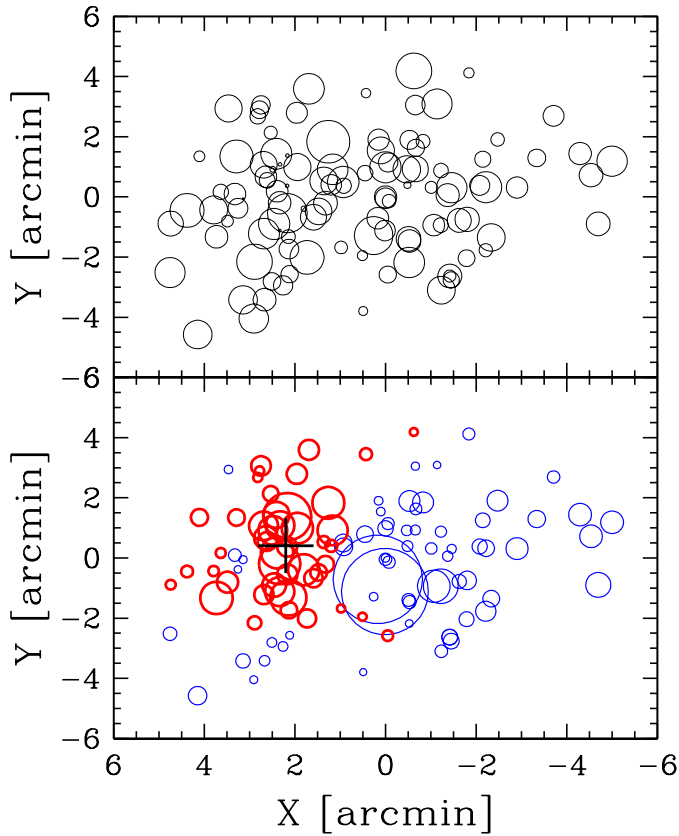}}
\caption
{ Spatial distribution of the 110 cluster members, each marked by a
  circle.  The plot is centered on the cluster center.  {\em Upper
    panel:} the cluster velocity field: the larger the circle, the
  larger is the galaxy velocity.  {\em Lower panel:} the result of the
  (modified) DS-test: the larger the circle, the larger is the
  deviation $\delta_{{\rm s},i}$ of the local velocity dispersion from the
  global velocity dispersion. Thin/blue and thick/red circles show
  where the local velocity dispersion is smaller or larger than the
  global value.  The cross indicates the position of the E-peak as
  detected in the 2D analysis (INT sample). }
\label{figdssegno10}
\end{figure}

\begin{figure}
\centering 
\resizebox{\hsize}{!}{\includegraphics{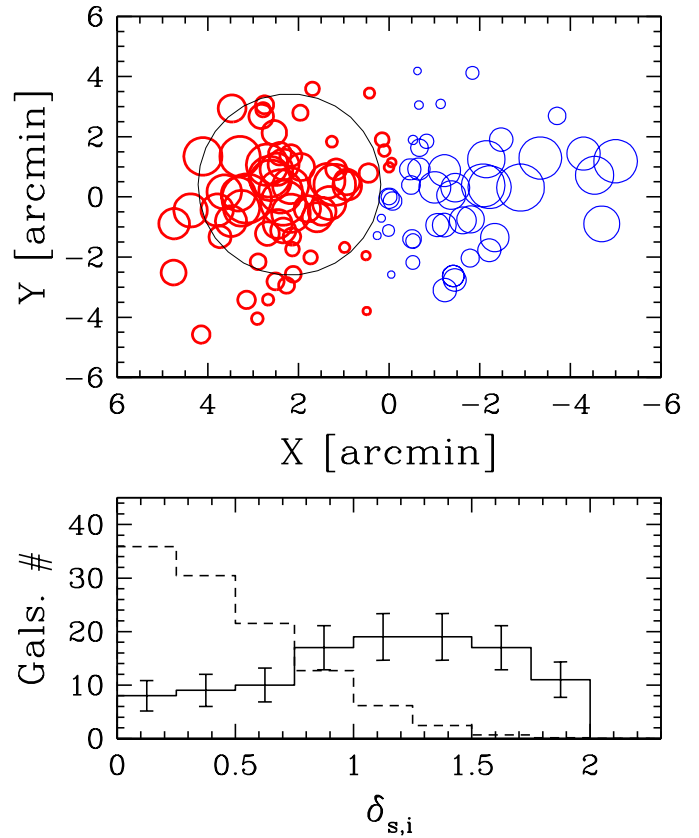}}
\caption
{{\em Upper panel:} the same of Fig.~\ref{figdssegno10} -- lower panel
  but for the DS-test with a larger number of neighbors
  ($N_{\rm{nn}}=40$). The large, faint circle encloses the region
  within 1.5\arcm from the E-peak: galaxies inside defines the
  E1.5-sample.  {\em Lower panel:} the distribution of $\delta_{{\rm s},i}$
  deviations of the above DS-test.  The solid line represents the
  observations, the dashed line the distribution for the galaxies of
  the 1000 simulated clusters, normalized to the observed number.  }
\label{figdssegno40}
\end{figure}

We also consider two kinematical estimators alternative to the
$\delta$ parameter of the DS-test, i.e. we consider separately the
contributes of the local mean $\delta_{\rm V}^2= [(N_{\rm
    nn}+1)/\sigma_{\rm V}^2](\overline {\rm V_l} - \overline {\rm
  V})^2]$, and dispersion $\delta_{\rm s}^2= [(N_{\rm
      nn}+1)/\sigma_{\rm V}^2](\sigma_{\rm V,l} - \sigma_{\rm V})^2]$
    (see, e.g.  Girardi et al. \cite{gir97}, Ferrari et
    al. \cite{fer03}).  When considering the $\mathrm{\delta_s}$
    estimator, we find evidence for peculiar local velocity dispersion
    at the $97\%$ c.l..  Figure~\ref{figdssegno10} -- lower panel --
    shows the distribution on the sky of all galaxies, each marked by
    a circle: the larger the circle, the larger the deviation
    $\delta_{{\rm s},i}$ of the local velocity dispersion from the
    global cluster value. The two largest deviations are obtained for
    the subgroups associated to two galaxies just south of the BCG for
    which $\sigma_{\rm{V},l}\sim 400$ \kss.  Other deviations interest
    the eastern region characterized by large values of the velocity
    dispersion. To better investigate we repeat the DS-test
    increasing the number of neighbors. As for the $\delta_{\rm s}$
    indicator, we obtain large c.l. values for peculiarity up to a
    $99.99\%$ c.l. in the $N_{\rm nn}$ =25--50 range (see
    Fig.~\ref{figdssegno40} -- upper panel).  This result suggests
    that the peculiarity interests a large part of the cluster. To
    individuate how many galaxies are involved in the substructure, we
    resort to the technique developed by Biviano et
    al. (\cite{biv02}), who used the individual $\delta_i$-values of
    the DS method.  The comparison of the $\delta_{{\rm s},i}$-values
    of all 1000 Monte Carlo simulations and the observed values shows
    how many galaxies are involved in the structures with peculiar
    kinematics (Fig.~\ref{figdssegno40} -- lower panel).

\subsubsection{Identifying subclusters in the eastern region}
\label{subclusters}

\begin{figure}
\centering 
\resizebox{\hsize}{!}{\includegraphics{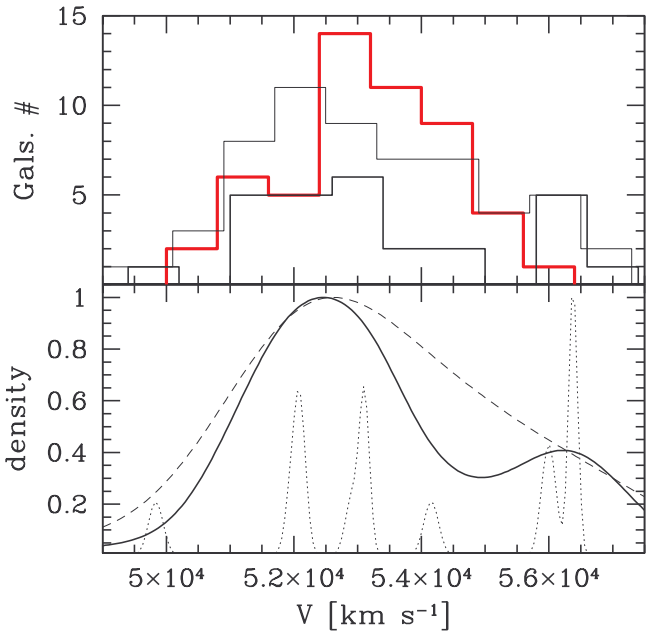}}
\caption
{{\em Upper panel:} velocity distributions of the galaxies in the
  western and eastern samples (thick/red and thin/black lines,
  respectively).  The velocity distribution of a subsample of the
  eastern sample, i.e. the 27 galaxies within a radius of 1.5\arcm
  from the E-peak (E1.5-sample), is also shown (black thick
  line). {\em Lower panel:} the galaxy density distribution in the
  velocity space, as provided by the DEDICA adaptive-kernel
  reconstruction method for the E1.5-sample (solid line) and for two
  samples contained within a larger and a smaller radii, E2.0- and
  E1.0-samples (dashed and dotted lines, respectively).  Unit on the
  y-axis is normalized to the density of the highest peak.}
\label{figdenEW}
\end{figure}

Comparing the two subsamples obtained by a rough division of A2254 (e.g., the
eastern and the western subsamples with respect to the position of the
BCG), we find that the eastern sample is characterized by a larger
value of $\sigma_{\rm{V}}$ (see Table~\ref{tabsub} and
Fig.~\ref{figdenEW} -- upper panel). The F-test confirms that the
velocity dispersions of the two samples differ at the 99.4\% c.l.; no
difference is shown between the mean velocities according to the
means-test (see,e.g., Press et al. \cite{pre92} for these tests).

Figure~\ref{figdssegno40} -- lower panel shows as the region with high
$\sigma_{\rm{V},l}$ values is centered roughly around the E-peak as
determined in the 2D analysis.  We select the galaxies within a
circular region centered in the position of the E-peak. Using a
radius of 1.5\arcmm, comparable to the distance between E- and
W-peaks, we obtain a subsample of 27 galaxies, the E1.5-sample,
characterized by a large value of $\sigma_{\rm{V}}\sim 1700$
\kss. When applying the 1D-DEDICA method to this subsample we detect
two significant peaks with 21 and 6 galaxies at $\sim$ 52\,500 and
56\,300 \kss, respectively (see E1.5LV and E1.5HV in
Table~\ref{tabsub}). Figure~\ref{figdenEW} -- lower panel -- shows the
results of the 1D-DEDICA method for the E1.5-sample, as well as for
other two samples defined by a larger and a smaller radius (2\arcm and
1\arcmm).  The choice of a large radius gives one-(asymmetric)
peak in the galaxy density distribution in the velocity space, while
a too small radius gives a list of non-significant peaks due to the poor
statistics involved.

\input{tabsub.tex}

\subsubsection{Identifying subclusters through the Htree method}
\label{subclustersh}

We resort to the method devised by Serna \& Gerbal (\cite{ser96},
hereafter Htree-method; see, e.g., Durret et al. \cite{dur10} for a
recent application). This method uses a hierarchical clustering
analysis to determine the relationship between galaxies according to
their relative binding energies.  The method assumes a constant value
for the mass-to-light ratio of galaxies and Serna \& Gerbal suggest
a value comparable to that of clusters.  Here we take a value of
$M/L_r$=150 \ml as suggested by large statistical studies (e.g.,
Girardi et al. \cite{gir00}; Popesso et al. \cite{pop05}).  The
(gross) results are quite robust against the choice of the value of
$M/L_r$.

\begin{figure}
\centering 
\resizebox{\hsize}{!}{\includegraphics{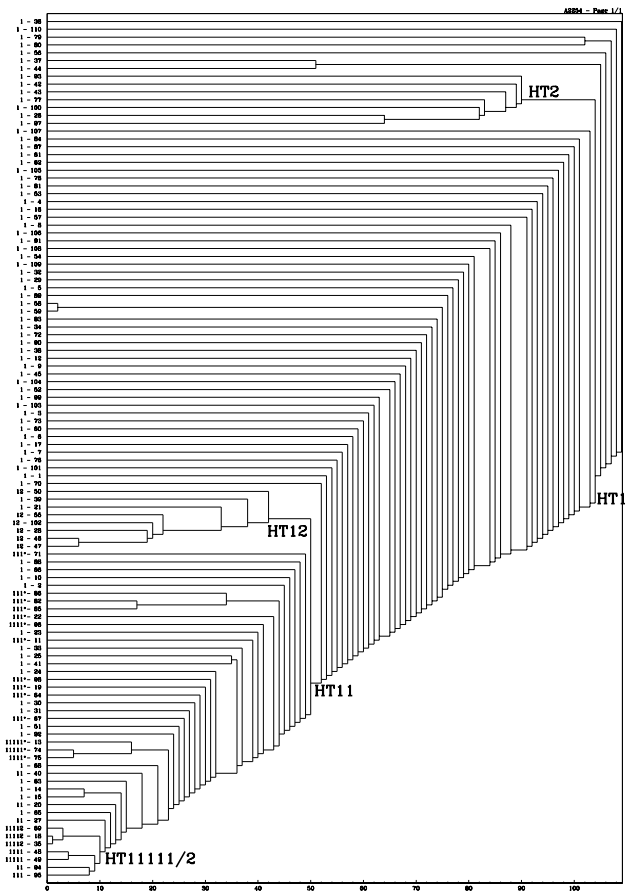}}
\caption
{Dendogram obtained through the Serna \& Gerbal (\cite{ser96})
  algorithm.  The abscissa is the binding energy (here in arbitrary
  unit with the deepest negative energy levels on the left) while the
  catalog numbers of the various member galaxies are shown along the
  ordinate (IDm  in Table~\ref{catalogA2254}).} 
\label{a2254gerbal}
\end{figure}

Figure~\ref{a2254gerbal} shows the resulting dendogram, where the
total energy appears horizontally.  Galaxy pairs and subgroups of
galaxies then appear with a lower total energy. Starting from the less
deep part of the energy levels (right part of the dendogram) we find
HT1, which we interpret as the main system, and the group HT2.  HT2 is
a high-velocity eastern group and contains all members of E1.5HV.
HT1 is a ``substructured'' main system and is formed by HT11, which
has the BCG in its potential well, and HT12, i.e. a few low velocity
galaxies located in the western region (in particular, the
BCG3). Note, however, that at the bottom of the binding energy of HT11
we find HT1111, which is not a core in the classical sense. In fact,
HT1111 contains HT11111 and HT11112. HT11112 is formed by three
galaxies close to the cluster center: the BCG and BCGc pair, and a
small compact galaxy located between them. HT11111 is close to the
E-peak and is formed by four galaxies, among which the BCG2 and BCG4.

\subsubsection{Identifying subclusters through the 3D-DEDICA method}
\label{subclustersd}

Finally, we apply the 3D-DEDICA method (Bardelli et
al. \cite{bar98}).  Table~\ref{tabsub} lists the five most significant
($\simg$ 94\% c.l.) groups (DED1--DED5) composed by at least three
galaxies.  The picture revealed by the 3D-DEDICA test is that of a
very complex structure, with three and two subclusters in the eastern
and western regions, respectively.

$\ $

\begin{figure}
\centering 
\resizebox{\hsize}{!}{\includegraphics{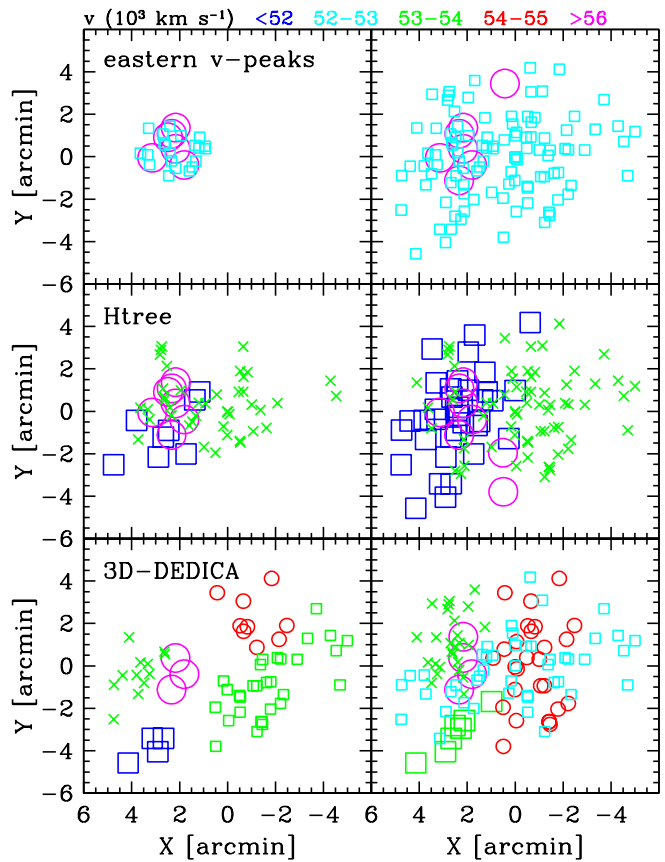}}
\caption
{Spatial distribution on the sky of the galaxies of 
the subclusters detected through our 3D analysis
and the cluster partitions obtained through the 3D-KMM
method using them as seeds ({\em left and right panels}, respectively).
In the {\em left panels} symbols refer to different groups.
The different colors refer to different mean velocities
according to the legend at the top of the figure.
In the {\em right panels} symbols indicate subclusters
found  using as seeds the respective groups on the left
and colors refer again to the velocity according to the above legend.
}
\label{figxy6}
\end{figure}

Figure~\ref{figxy6} -- left col. -- shows the spatial distribution of
the groups detected by above methods and Table~\ref{tabsub} lists the
kinematical properties: the number of assigned members, $N_{\rm g}$
(Col.~2); the mean velocity, $\left<\rm{v}\right>$ (Col.~3); the
velocity dispersion with bootstrap errors, $\sigma_{\rm V}$ (Col.~4);
a short comment summarizing our interpretation (Col.~5). Following
properties refer to the corresponding KMM subclusters, see below
(Cols.~6, 7, and 8).

\subsubsection{3D-KMM method}
\label{2dkmm}

We use the above results to apply the full 3D-KMM method, which,
starting from the above seeds, allow us to assign all galaxies to
subclusters.  

Using E1.5LV and E1.5HV as seeds for a initial two-group partition, we
find that a partition of 103 and 7 galaxies is a significantly more
accurate description of the 3D galaxy distribution than a single 3D
Gaussian (at the 99.7$\%$ c.l.). We note that the KMM-E1.5LV
subcluster. i.e. the main system, has a significant weighted gap in
its velocity distribution (as in the whole sample, see
Sect.~\ref{velo}) but does not show any trace of peculiar kinematics
according to the DS-test.  Using HT1, HT2, and HT1111 as initial
seeds, the 3D-KMM converges to a three-group partition significant at
the 97.5$\%$ c.l..  Finally, using DED1--DED5 groups as initial seeds
we find the 3D-KMM test converges to a five-group partition
significant at the 99.9$\%$ c.l..

Figure~\ref{figxy6} -- right col. -- shows the spatial distribution of
the obtained partitions.

\section{X-ray morphological and spectral analysis}

\subsection {Observation and data reduction}
\label{xray}

A2254 was observed by \xmm\ on August 26, 2009. The data were reduced
with SAS v10.0 using the tasks {\em emchain} and {\em epchain}. We
only considered event patterns 0--12 for MOS and 0 for pn, and the
data were cleaned using the standard procedures for bright pixels and
hot columns removal (by applying the expression FLAG == 0) and pn
out-of-time correction. The energy scale of the pn over the whole
spectral bandpass was further improved with the task {\em epreject}.
Periods of high background event counts due to soft protons were
filtered using a threshold cut method: we performed a Gaussian fit to
the peak of the histogram of the light curve and excluded the time
bins where the count rate lies more than 3$\sigma$ away from the
mean. The light curves were extracted from regions of least source
contamination (excising the bright object core in the central
8\arcmin\ and the point source list from the SOC pipeline, after
visual inspection) in two different energy bands: a hard band, 10--12
keV for MOS and 10--13 keV for pn (using 100s bins), and a wider band,
0.5--10 keV (using 10s bins), as a safety check for possible flares
with soft spectra. The flaring periods thus determined were further
checked by visual inspection of the light curves, resulting in net
exposures of 57.0 ks, 56.8 ks, and 45.1 ks for the MOS1, MOS2, and pn
detectors, respectively. Point sources were detected using the task
{\em ewavelet} in the energy band 0.5--10 keV and checked by eye on
images generated for each detector. Detected point sources from all
detectors were merged, and the events in the corresponding regions
removed from the event list using circular regions of
25\arcsec\ radius centered at the source position. The area lost by
point source exclusion, CCD gaps, and bad pixels was calculated using
a mask image. Redistribution matrix files (RMFs) and ancillary
response files (ARFs) were generated with the SAS tasks {\em rmfgen}
and {\em arfgen}, the latter in extended source mode. Appropriate
flux-weighting was performed for RMFs, and for ARFs using
exposure-corrected images of the source as detector maps (with pixel
size of 1\arcmin, the minimum scale modeled by {\em arfgen}) to sample
the variation in emission, following the prescription of Saxton \&
Siddiqui (\cite{sax02}). For the background subtraction, we adopted a
complete modeling of the various background components as described in
detail in Gastaldello et al. (\cite{gas07}).  We accurately modeled
the actual sky background using a region free of cluster emission at
$r>9$\arcmin, where the surface brightness profile in the 0.7--1.2 keV
reaches the background level.  The spectra of the
out-of-field-of-view events of CCD 4 and 5 of MOS1 showed an
anomalously high flux in the soft band (see Kuntz \& Snowden
\cite{kun08}) so they were excluded from our surface brightness and
spectral analysis.

\subsection {X-ray image and surface brightness analysis}
\label{ximage}

\begin{figure}
\centering
\resizebox{\hsize}{!}{\includegraphics{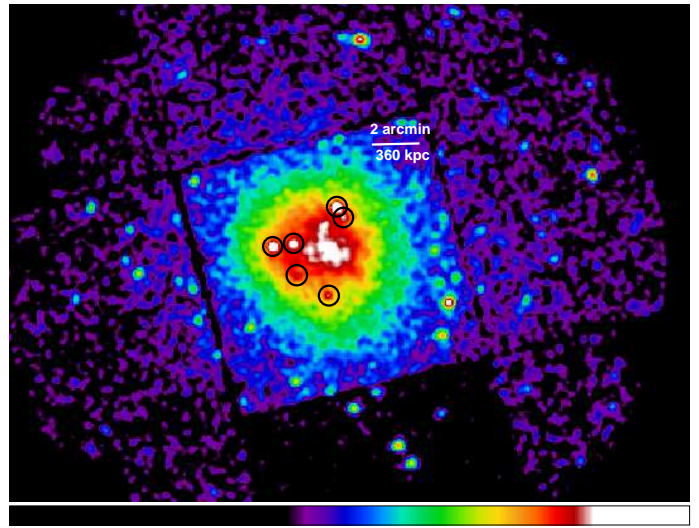}}
\caption
{Mosaic of the MOS1 and MOS2 images in the 0.5--2 keV energy band
  smoothed on a 12\arcsec scale. The image was divided by the summed
  exposure maps to correct for exposure variations. The detected point
  sources within the central regions of the cluster discussed in the
  text are highlighted by black circles.}
\label{fig:imax}
\end{figure}

\begin{figure}
\centering
\resizebox{\hsize}{!}{\includegraphics{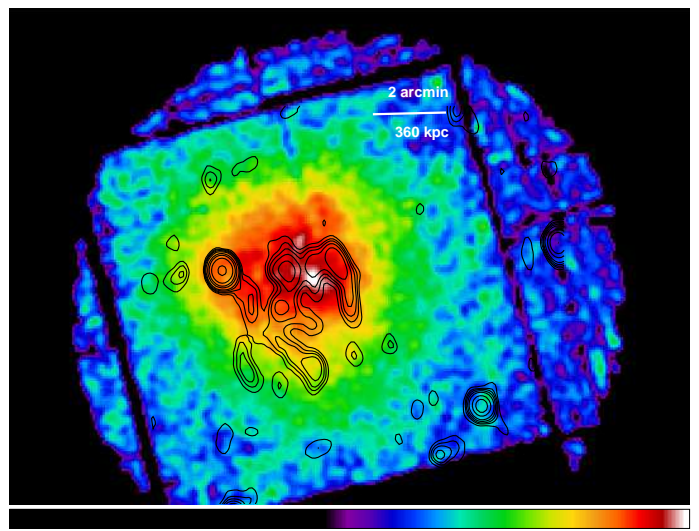}}
\caption
{X-ray image of the diffuse emission of the A2254 cluster where point
  sources were removed and replaced by a local estimate of the
  surrounding emission and only regions exposed by more than 65\% of
  the maximum exposure is considered. The contour levels of the radio
VLA image (Govoni et al. \cite{gov01b}) are superimposed, too}
\label{fig:imax_nopoint}
\end{figure}

For each MOS detector, we create images in the 0.5--2 keV band and
exposure maps at the monochromatic energy of 1 keV and then combine
the images into a single exposure-corrected image, smoothed on a scale
of 12\arcss, as shown in Fig.\ref{fig:imax}. We process the image to
remove the point sources using the CIAO tool {\em dmfilth}, which
replaces photons in the vicinity of each point source with a locally
estimated background. We consider only the regions exposed by more
than 65\% of the maximum exposure (see Fig.~\ref{fig:imax_nopoint}) in
order to perform the X-ray morphology analysis. We adopt quantitative
measures by applying the power ratios technique (Buote \& Tsai
\cite{buo96}), the centroid shifts (Mohr et al. \cite{moh93}) and the
concentration parameter (Santos et al. \cite{san08}).

Power ratios derive from multipole expansions of the two-dimensional
projected gravitational potential and they can be applied directly on
X-ray images since under normal circumstances the X-ray emissivity
can be used as a proxy for the gravitational potential. Power ratios
are computed within a fixed circular aperture and they can be computed
on the centroid of the cluster or on the X-ray peak. In the former
case $\rm{P_1/P_0}$ is null, $\rm{P_2/P_0}$ gives information on the
ellipticity of the cluster, $\rm{P_3/P_0}$ is related to a bimodal
structure and $\rm{P_4/P_0}$ is sensitive on small scale
substructure. In the latter case $\rm{P^{pk}_1/P^{pk}_0}$ is non null
if the cluster does not exhibit reflection symmetry about two
orthogonal axes originating on the peak of the emission and it is
essentially a circularly average centroid shift. We calculate the
power ratios within an aperture of 500 kpc, given the high
signal-to-noise within this radius and for comparison with previous
work.

Centroid shifts indicate that the center of mass of the X-ray emitting
gas varies with radius.  Centroid shifts and power ratios are both
capable of identifying highly disturbed systems or systems with
significant, well defined substructures (Poole et
al. \cite{poo06}). However the centroid shifts parameter, $w$, is more
sensitive to subtler disturbances.  Following the method of O'Hara et
al. (\cite{oha06}, see also Poole et al. \cite{poo06}), the centroid
shift is computed in a series of circular apertures centered on the
cluster X-ray peak. The radius of the apertures is decreased in steps
of 5\% from 500 kpc to 25 kpc.

The concentration parameter, $c$, is defined as the ratio of the peak
(calculated within 100 kpc) over the ambient surface brightness
(calculated within 500 kpc). We apply these techniques to the image
reported in Fig.~\ref{fig:imax_nopoint} and the results are reported
in Table~\ref{tabxmorpho}.

\input{tabxmorpho}

\begin{figure}
\centering
\resizebox{\hsize}{!}{\includegraphics[angle=-90]{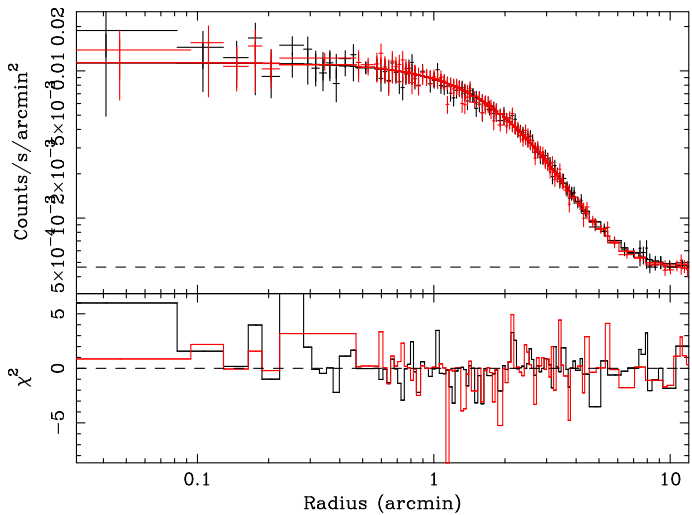}}
\caption
{Surface brightness profile of the X-ray emission of A2254.  Data from
  MOS1 and MOS2 are plotted in black and red, respectively.  The best
  fit $\beta$-model and ratio of data over the model are also shown. The
  background level is shown as the black dashed line in the lower
  panel.}
\label{fig:sbprofile}
\end{figure}

Following Ettori et al. (\cite{ett10}) we extract the surface brightness
profile in the energy band $0.7$--$1.2$ keV, in order to keep the
background as low as possible with respect to the source. For this
reason, we avoid the intense fluorescent instrumental lines of Al
($\sim 1.5$ keV) and Si ($\sim 1.8$ keV). We account for the X-ray
background by including a constant background component.  The data
are grouped to have at least 20 counts per bin in order to apply
the $\chi^{2}$ statistics.  The fitted model is convolved with the
\xmm\ PSF. The joint best-fit $\beta$-model has a core radius of $r_c
= (507\pm36)$ \kpc ($169$\arcsec$\pm12$\arcsec) and $\beta=0.95\pm0.06$
for a $\chi^{2}$/d.o.f. = 215/179 (see Fig.\ref{fig:sbprofile}).

\subsection {Spectral analysis}
\label{xspec}

\begin{figure}
\centering
\includegraphics[width=8cm, angle=-90]{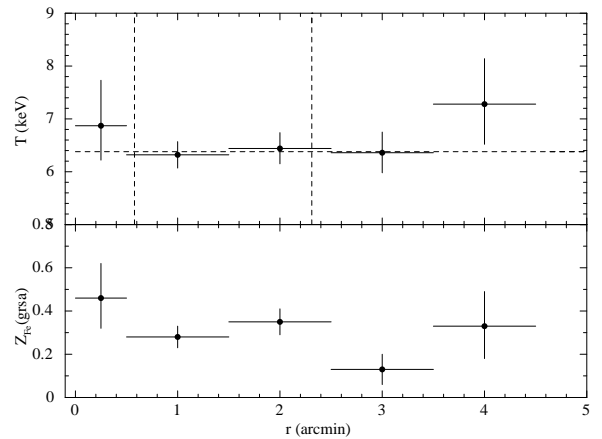}
\caption
{Profiles of the azimuthally averaged, projected temperature and metal
  abundances of A2254. The error bars are at the 68\% c.l..  The
    horizontal line shows the value of $kT_{\mathrm OUT}$, i.e.  the
    value of $kT$ estimated within the $0.05 R_{180} < r < 0.2
    R_{180}$ region, the two radii being indicated by the two vertical
    lines.}
\label{fig:tzprofile}
\end{figure}

We extract spectra for each detector in 5 concentric circular annuli
located at the the peak of the X-ray emission with bounding radii
0\arcmin--0.5\arcmin, 0.5\arcmin--1.5\arcmin, 1.5\arcmin--2.5\arcmin,
2.5\arcmin--3.5\arcmin, 3.5\arcmin--4.5\arcmin.  The spectra from the
three detectors are re-binned to ensure a signal-to-noise ratio of at
least 3 and a minimum 20 counts per bin (necessary for the validity of
the $\chi^{2}$ minimization method).  We fit the spectra with an APEC
thermal plasma modified by Galactic absorption fixed at
$4.7\times10^{20}$ cm$^{-2}$ (Kalberla et al.  \cite{kal05}) to each
annulus.  A correction file derived from the background modelling is
used in XSPEC as in Gastaldello et al. (\cite{gas07}).  The free
parameters are temperature, normalization (proportional to emission
measure), and the metallicity, with the solar abundance units of
Grevesse \& Sauval (\cite{grev98}).  The spectra are adequately fitted
by the model, with reduced $\chi^2$ ranging from 1.06 in the inner
annulus to 1.5 in the outermost annulus.  The azimuthally averaged,
projected temperature and abundance profile thus obtained are shown in
Fig.~\ref{fig:tzprofile}. The temperature profile is rather flat and
in particular there is no evidence of a decrease in the core, as found
in non-cool core clusters.  We calculate the mean temperature of the
cluster as the temperature obtained with a spectral fit in the region
$0.05 R_{180} < r < 0.2 R_{180}$, where $R_{180} = 1780 (kT /
5{\rm{keV}})^{1/2} h(z)^{-1}$ \kpc (Arnaud et al. \cite{arn05}, with $
h(z)= (\Omega_{\rm{M}} (1+z)^3 + \Omega_{\Lambda})^{1/2}$) using an
iterative procedure to calculate $kT$ and $R_{180}$ (Rossetti \&
Molendi \cite{ros10}).
We find $kT=
6.38 \pm 0.25$ keV
and $R_{180}=2.08$ \hh. The value of $kT$ thus estimated within the
  $0.05 R_{180} < r < 0.2 R_{180}$ region, hereafter $kT_{\mathrm
    OUT}$, is a good proxy for the global temperature (see Fig.~4 of
  Leccardi \& Molendi \cite{lec08} where $kT_{\rm M}$ is the global
  temperature). In particular as for A2254, Figure~\ref{fig:tzprofile}
  shows as the measure of $kT$ has already large uncertainties around 
  $0.4R_{180}$.

To characterize temperature and emission measure gradients, we
  compare the central to a global value of the pseudo-entropy measure.
Following Leccardi et al. (\cite{lec10}), we compute the
pseudo-entropy ratio ${\cal E }$ $= (kT_\mathrm{IN}/kT_\mathrm{OUT})
\times (E\!M_\mathrm{IN}/E\!M_\mathrm{OUT})^{-1/3}$, where $kT$ and
$E\!M$ are the X-ray temperature and emission measure, and
$\mathrm{IN}$ and $\mathrm{OUT}$ define the region within the radius
$\approx$0.05~$R_{180}$ and the region within the the annulus with
bounding radii 0.05-0.20~$R_{180}$, respectively.  Note that
  Rossetti \& Molendi (\cite{ros10}) proved that the pseudo-entropy
  ratio used in their study is a good proxy of the deprojected
  quantity and Rossetti et al. (\cite{ros11}, their Fig.~3) found a
  good correlation between the pseudo-entropy ratio and, for example,
  the central entropy defined in Cavagnolo et al. (\cite{cav09}).  We
measure ${\cal E}$ $= 0.833 \pm 0.097$ which places A2254 among
clusters with high entropy core  (see also Sect.~\ref{stru}).

\begin{figure}
\centering
\includegraphics[width=11cm,angle=90]{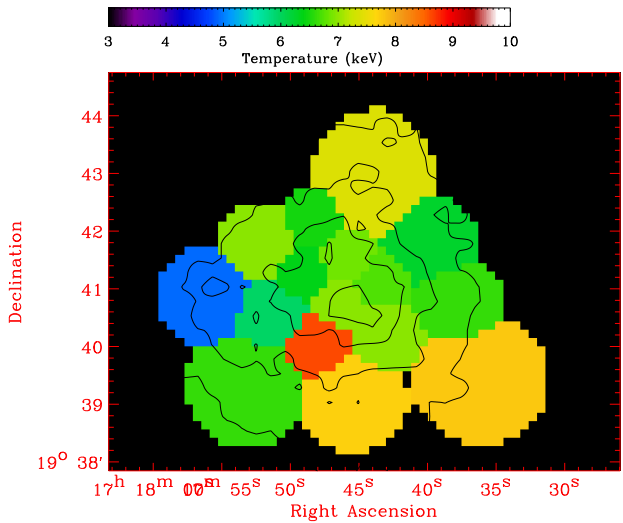}
\caption
{Temperature map (keV) with, superimposed, the contour levels of
    the XMM image. Coordinates are right ascension and declination
    (J2000).
}
\label{tmap}
\end{figure}

We also prepare a two dimensional map of X-ray temperature, starting
from the EPIC images. We use a modified version of the adaptive
binning and broad band fitting technique described in Rossetti et
al. (\cite{ros07}), where we substitute the Cappellari \& Copin
(\cite{cap03}) adaptive binning algorithm with the weighted Voronoi
tessellation by Diehl \& Statler (\cite{die06}, see Rossetti \&
Molendi \cite{ros10}).  The temperature map is shown in
Fig.~\ref{tmap}. The formal errors given in output from the procedure
are of the order of 15--$20\%$ but the measures at the map periphery, more
affected by the background, should be considered with more caution.

\section{Discussion and conclusions}
\label{disc}

\subsection{Cluster structure}
\label{stru}

The high values of the velocity dispersion 
$\sigma_{\rm  V}=1340_{-84}^{+101}$ \ks and X-ray temperature $kT=(6.38\pm0.25)$
keV are comparable to the values found for hot, massive clusters
(Mushotzky \& Scharf \cite{mus97}; Girardi \& Mezzetti \cite{gir01}).
However, our estimates of $\sigma_{\rm V}$ and $T_{\rm X}$ are
$3\sigma$-inconsistent each other under the assumption of the
equipartition of energy density between ICM and galaxies. In fact, we
obtain $\beta_{\rm spec} =1.71^{+0.26}_{-0.21}$ to be compared with
$\beta_{\rm spec}=1$, where $\beta_{\rm spec}=\sigma_{\rm v}^2/(kT/\mu
m_{\rm p})$ with $\mu=0.58$ the mean molecular weight and $m_{\rm p}$
the proton mass (see also Fig.~\ref{figprof}).  Evidence that A2254 is
well far from the dynamical equilibrium comes from both optical and
X-ray analyses as discussed in the following.

From the optical point of view, we find a strong evidence of
bimodality from our 2D analysis of the photometric sample with the E
and W-peaks far by $\sim$0.4--0.5 \hh. The (modified) DS-test also shows
evidence of substructure at a very high level (at the $99.99\%$ c.l.).
Our 3D analysis of the region of the 2D E-peak, detects a high
velocity subcluster which explains the high velocity dispersion in
that region and the strong increase of velocity dispersion profile at
$\sim 0.4$--0.5 \h from the cluster center (see Fig.~\ref{figprof}).

However, a simple bimodal structure is not completely satisfactory
when looking at the whole observational scenario. In fact, the W-peak
is not well centered onto the BCG and both E- and W-peaks seem to
have, at their turn, a bimodal or a N-S elongated structure (see
Figs.~\ref{figk2zoom} and ~\ref{figVTP}).  The E-subcluster recovered
through the 3D analysis is a minor structure and contains no dominant
galaxy: it seems improbable that it causes the presence of the 2D
E-peak which is almost as rich as the W-peak (see
Table~\ref{tabdedica2d}).  Indeed, more refined 3D analyses (e.g., the
H-tree and the 3D-DEDICA methods) show evidence for a more complex
structure. In particular, the H-tree method detects a low-velocity
eastern subcluster which can explain even better the important 2D
E-peak and the strong increase of the velocity dispersion profile at
$\sim 0.4$--0.5 \h from the cluster center.  The low-velocity eastern
subcluster might also explain the results of the weighted gap analysis
(see Fig.~\ref{figstrip}). The 3D-DEDICA analysis reveals five
subclusters, two of which in the western region: they might explain
the presence of a bimodal 2D W-peak.

We stress that our identification of subclusters are always confirmed
by the 3D-KMM analysis. In principle, one might use the significance
(as computed according to the likelihood ratio test statistics), to
select the best cluster partition (e.g., Bird \cite{bir94}).  However,
we note that in A2254, the high velocity E-group is the only
subcluster well contained in the region sampled by our redshift data,
while other subclusters have a large size and/or are at the boundary
of this region.  Thus, although the use of the 3D-KMM approach seems
useful to perform a homogeneous check of the subclusters detected
using other disparate methods, we do not rely on the 3D-KMM results to
fix the best partition of subclusters and the relevant kinematical
properties.  To obtain more stringent conclusions on the structure of
A2254 we stress the need of more galaxy redshifts and a more extended
sampled region.

As for the X-ray data, all the methods of the morphological analysis
provide evidence of a dinamically disturbed cluster: i) the power
ratio analysis, with its value of $\rm{P^{pk}_1/P^{pk}_0}=1.85\pm
0.47$ together with the radio power at 1.4 GHz of $P_{\rm
  1.4GHz}=2.9\times10^{24}\ h_{70}^{-2}$ W\ Hz$^{-1}$ (Govoni et
al. \cite{gov01b}) places A2554 in the region of well known radio halo
mergers, such as Abell 2256 and Abell 665, on the $P_1/P_0$ -- $P_{\rm
  1.4GHz}$ relation of Buote (\cite{buo01}; see their Fig.~1-left,
where $P_{\rm 1.4GHz}=2.1\times10^{24}$ W\ Hz$^{-1}$ in their
cosmology); ii) the combination of the three morphology indicators (as
done by Cassano et al. \cite{cas10b}) in the $(w,c)$, $(P_3/P_0,c)$,
and $(P_3/P_0,c)$ planes designates A2254 as a morphologically
distubed cluster among the other clusters with radio halo in the
sample of that paper. In fact, A2254 shows well higher values of $w$
and $P_3/P_0$ and a lower value of $c$ with respect to the median
discriminating values between disturbed and relaxed systems,
i.e. $w=0.012\times 500$ \kpcc, $c=0.2$, and $P_3/P_0= 1.2\times
10^{-7}$. iii) The value of $w=(0.037\pm0.001)\times 500$ \kpc gives a
significant indication of a non relaxed system. It is an intermediate
value among those measured for 45 ROSAT-observed clusters by O'Hara et
al. (\cite{oha06}). According to the simulations by Poole et
al. (\cite{poo06}, their Fig.~11) this value of $w$ is typical of
ongoing cluster mergers.

The active dynamical state of A2254 is also confirmed from the
thermodinamic point of view, given the absence of a temperature drop
in the core and the classification of A2254 as an high entropy core
according to Leccardi et al. (\cite{lec10}) who fixed the
  separation value between medium and high entropy core cluster at
  ${\cal E}$=0.64 (${\cal E}$= $0.833 \pm 0.097$ in A2254).  In fact,
  observations seem to favor cool core destruction through cluster
  mergers (Allen et al. \cite{all01}; Sanderson et al. \cite{san06})
  and the vast majority of merging systems hosts high entropy cores
  according to Leccardi et al. (\cite{lec10}), who analyzed 59 clusters
  spanning a large range of dynamical states.  The temperature map,
  limited by the quality of the data to the inner regions, reveal a
  hot core with no significant temperature variations, consistent with
  an on-going merger.

\subsection{Optical -- X-ray -- radio data comparison}
\label{cfr}

At large scale, when comparing data from different wavelengths,
  the main feature is the correlation between the elongation of the
  optical structure, with the two galaxy clumps far by $\sim$0.5 \h
  along the E-W direction, and the E-W elongation of the X--ray
  emission isophotes (see Fig.~\ref{figimage}). The elongation of
  X-ray isophotes in the merging direction is expected from
  simulations (e.g., Roettiger et al. \cite{roe96}).
At small scale, we note that
the X-ray emission is peaked at the BCG galaxy. Instead, the E-peak
does not coincide with a secondary X-ray peak.  This discrepancy
between the distributions of the ICM (collisional) and galaxies (non
collisional component) is a strong point in favor of an ongoing
cluster merger.

Our interpretation of the above correlations or no-correlations 
in the context of A2254 scenario is
  the following.  The almost ``triangular'' shape of external X-ray
isophotes indicates that the high velocity E-group has just passed
through the main system going from west to east, as suggested by the
comparison of Fig.~\ref{fig:imax_nopoint} with, e.g., the second or
third panels -- or maybe the fourth panel considering a possible
projection effect -- in Figs.~4 and ~5 of Poole et al.~(\cite{poo06}
-- top line, where the subcluster comes from the left). This scenario
agrees with the fact that we observe the high-velocity E-group peak of
galaxies being more advanced in the merging direction while the ICM
component is slowed down and left behind (see e.g., the bullet cluster
by Markevitch et al. \cite{mar02}).  At difference with the bullet
cluster, A2254 does not reveals a secondary X-ray peak: this might be
due to a more advanced phase of the merger or, in agreement with our
mass estimates (see below), with the fact that the subcluster is much
smaller and thus completely stripped by its gas content. 

However, a simple bimodal merger is not completely satisfactory.
In fact, the X-ray emission in the internal cluster region, 
i.e. in the region of the W-peak, also reveals an elongation toward the
North, perpendicular to the merging direction (see
Figs.~\ref{fig:imax} and \ref{fig:imax_nopoint}). This could be
connected to a particular phase of the the E-W merger resembling the
high emission structure perpendicular to the merger at the time of the
apocenter (but not in its symmetric aspect, see Fig.~4 of Poole et
al.~\cite{poo06}). Alternatively, it could be connected with another
accretion phenomenon -- along the N-S direction. In fact, optical
  2D analyses of deep Subaru data (see Figs.~\ref{figk2zoom} and
  Figs.~\ref{figVTP}) reveal a N-S double structure in the cluster
  center: one enclosing the BCG and one slightly at north. The 3D DEDICA
  analysis finds  a northern and a southern galaxy clumps in the western
  part of the cluster, too.

As for its radio morphology, A2254 is the most irregular in its radio
morphology in the sample of Govoni et al. (\cite{gov01b}).  The halo
is not centered on the region of the maximum X-ray emission (see its
upside-down ``U'' shape in Figs.~\ref{figimage} and
\ref{fig:imax_nopoint}), resembling in some respects the behavior of
radio relics.  As for the comparison between radio and other
  data, the main feature is its bad correlation, already at large
scales,  with X-ray emission,
  and with the merging direction suggested by the two optical
  subclusters.  In fact, here the halo does not permeate the cluster
  volume in a similar way to the X-ray emitting gas of the
  intracluster medium (see Fig.~\ref{figimage}) as generally shown in
  other clusters not only by their large scale morphology but also by
  the point-to-point correlation between radio and X-ray brightness
  (see the case of the four clusters analyzed by Govoni et
  al. \cite{gov01a}).  Moreover, halos are round or elongated in the
direction of the merger (the bullet cluster by Markevitch et
al. \cite{mar02}; Abell 520 by Girardi et al. \cite{gir08}; Abell 754
by Macario et al. \cite{mac11}): this is not the case of the A2254
  halo. 

Our interpretation is that the peculiar radio morphology of
  A2254 is likely due to the very complex substructure of A2254.  In
particular, the two N-S elongated branches of the radio emission
  might be connected with the minor N-S mergers suggested by our
  optical analysis.  However, note that, although the ``U'' shape
  morphology of the radio halo (see Fig.~\ref{figimage}) recalls the
  ``H' shape of the contours in the galaxy distribution in
  Fig.~\ref{figk2zoom}, the two features are shifted one respect to
  the other along the E-W direction.

Finally, we stress the absence of a connection between bright
  galaxies and subclusters, while this is often found in complex
  systems in 2D (e.g., Geller \& Beers \cite{gel82}), in 3D (e.g.,
  Abell 520, Girardi et al. \cite{gir08}), and also when a subcluster
  is partially destroyed (see the beautiful example of the tidal
  debris in A2744 by Owers et al. \cite{owe11}). In A2254, neither the
  high-velocity E-peak, the best detected through our optical
  analysis, hosts a bright, dominant galaxy. The feature which is
  worthy of mention, in spite of the poor statistics, is that BCG2,
  BCG3, and BCG4 are all located on the eastern side of the cluster,
  more external than respect to the E-subcluster. However, their
  optical spectra do not show features -- e.g. emission lines or a
  strong H$_{\delta}$ absorption -- typical of a possible starburst or
  poststarburst activity triggered by the shocked gas during a
  cluster-cluster merger (Bekki et al. \cite{bek10}), thus there is no
clear connection with the ongoing merger.

\subsection{A simple model for mass and merging scenario}
\label{mass}

Following the above discussion, A2254 can be described in first
approximation by a main system and an eastern high velocity
group. Looking at Table~\ref{tabsub} we can estimate $\sigma_{\rm
  V}\sim$ 1000--1200 \ks and $\sigma_{\rm V}= 200$--500 \ks for the
main and the E-group, respectively.

Making the usual assumptions for the main system and the subcluster
(cluster sphericity, dynamical equilibrium, coincidence of the galaxy
and mass distributions), we can compute virial global quantities.  We
follow the prescriptions of Girardi \& Mezzetti (\cite{gir01}, see
also Girardi et al. \cite{gir98}) and compute $R_{\rm vir}$ -- an
estimate for $R_{\rm 200}$ -- and the mass contained within this
radius.  In particular, we assume for the radius of the
quasi-virialized region $R_{\rm vir}=0.17\times \sigma_{\rm v}/H(z)$
\hh (see Eq.~1 of Girardi \& Mezzetti \cite{gir01} with the scaling of
$H(z)$ of Eq.~ 8 of Carlberg et al. \cite{car97} for $R_{200}$).  For
the mass we use $M=M_{\rm V}-SPT=3\pi/2 \cdot \sigma_{\rm V}^2 R_{\rm
  PV}/G-{\rm SPT}$ (Eq.~3 of Girardi \& Mezzetti \cite{gir01}), with
${\rm R}_{\rm PV}$ computed using the full procedure (see also Eq.~13
of Girardi et al. \cite{gir98}, with $A=R_{\rm vir}$ for a close
approximation) and the surface pressure term correction $SPT=0.2\times
M_{\rm V}$.  In practice, both $R_{\rm vir}$ and $M$ can be computed on
the basis of the estimated velocity dispersion with the usual
  scaling-laws $R_{\rm vir} \propto \sigma_{\rm V}$ and
  $M(<R_{\rm vir}) \propto \sigma_{\rm V}^3$.

We compute $M(<R_{\rm vir}=2.2$--2.7$ \hhh)=1.5$--2.5 \mqui and
$M(<R_{\rm vir}=0.4$--1.2$ \hhh)=0.1$--1.8 \mqua for the main and the
E-group respectively, leading to a small mass ratio for the merger
($<(1:10)$). We compute a mass $M_{\rm sys}(<R=2.2$--2.7$
\hhh)=1.5$--2.9 \mqui for the whole system, where we assume that
$M\propto R$ to roughly extrapolate the mass of the E-group.  
  According to the virial scaling laws, the above mass range leads to
  $\sigma_{\rm V}\sim (1000$--$1250)$ \ks for the virial
  velocity dispersion, thus making acceptable the $\beta_{\rm spec}=1$
  model.

For comparison, we would obtain $M_{\rm sys,1}(<R_{\rm vir,1}=3.0
\hhh)=(3.5\pm1.0)$ \mqui when assuming A2254 as a dynamically relaxed
cluster and $M_{\rm sys,5}\sim 4$ \mqui when assuming A2254 is the
very complex cluster formed by the five DEDICA subclusters.

We use X-ray data to compute an alternative mass estimate.  Given
  the no significant departures from the isothermality of the
  temperature profile out to $\sim$ 0.9 \h (see
  Fig.~\ref{fig:tzprofile}), we derive a mass estimate using the
  isothermal beta model where the total mass profile can be expressed
  by a simple analytical formula (e.g., eq. 6 of Henry et
  al. \cite{hen93} with $\gamma=1$).  Using the value of $kT_{\rm
    OUT}$ and the fitted parameters of the $\beta$ model to the
  surface brightness (Sect.~\ref{xray}), we obtain a mass $M_{\mathrm
    X}(<R=0.9 \hhh)=0.48$ \mqui with nominal errors less than the
  $5\%$.  We rescale the optical mass $M_{\rm sys}$ to $R=0.9$ \h
  assuming that the system is described by a King-like mass
  distribution with a very small core radius (Girardi et
  al. \cite{gir98}) or, alternatively, a NFW profile where the
  mass-dependent concentration parameter $c$ is taken from Navarro et
  al. (\cite{nav97}) and rescaled by the factor $1+z$ (Bullock et
  al. \cite{bul01}; Dolag et al. \cite{dol04}), here $c\sim 4$. We
  obtain $M_{\mathrm{sys}}(<R=0.9 \hhh) =(0.6$--$1.0)$ \mqui to be
  compared with the above $M_{\mathrm X}(<R=0.9 \hhh)\sim 0.5$
  \mquii. Thus, when opportunely correcting the optical mass for the
  presence of substructure, we do not find a large discrepancy between
  optical and X-ray mass estimates in agreement with that $\beta_{\rm
    spec}=1$ is acceptable (see above).

Taking the above face on values, one might estimate that the
non-thermal pressure in A2254 ranges from a value of $10\%$ to $100\%$
of the gas thermal pressure.  However, note that the above comparison
of the two mass estimates should be considered at best a tentative
since the multiclump structure of the cluster should be better
known. Indeed, previous efforts on making a comparison to put a limit
on non-thermal pressure have targeted systems which can be assumed in
equilibrium at least in the non-collisional component (e.g., a sample
of elliptical galaxies by Churazov et al. \cite{chu10}).

We can use our results to examine A2254 with respect to the observed
scaling relations among $P_{\rm 1.4GHz}$, the halo radio power, and
$M_{\rm H}$, the total cluster mass contained within the size of the
radio halo, $R_{\rm H}$ (Cassano et al. \cite{cas07}). We rescale the
mass $M_{\rm sys}$ to $R_{\rm H}=0.4$ \h (Cassano et al. \cite{cas07})
and obtain $M_{\mathrm{sys}}(<R_{\rm H}=0.4
\hhh)=(0.2$--$0.5)$ \mquii. Thus A2254 is well in agreement with the
relation fitted for other clusters with radio halos (see Fig.~9 of
Cassano et al. \cite{cas07}).

\begin{figure}
\centering
\resizebox{\hsize}{!}{\includegraphics{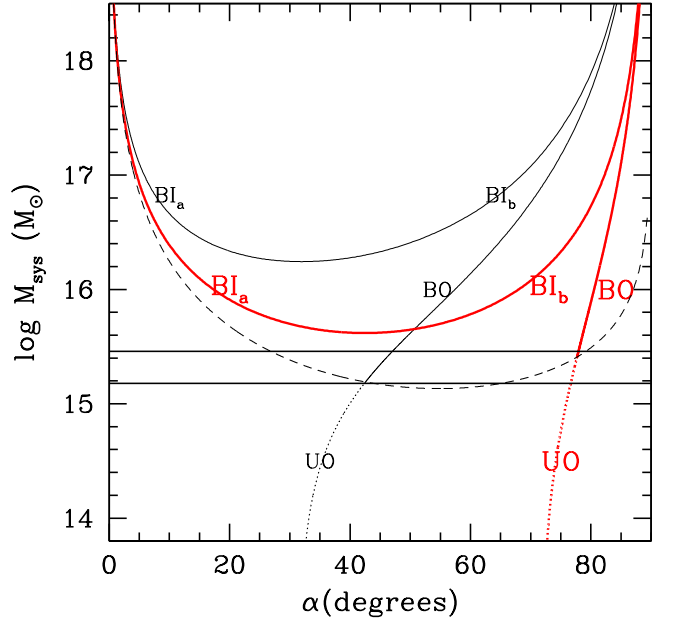}}
\caption
{System mass vs. projection angle for bound and unbound solutions
  (thick solid and thick dashed curves, respectively) of the two--body
  model applied to the high velocity E-group and the main cluster.
  Thin/black and thick/red lines refer to $t=0.1$ Gyr and $t=0.5$ Gyr,
  respectively.  Labels BI$_{\rm a}$ and BI$_{\rm b}$ indicate the
  bound and incoming, i.e., collapsing solutions (solid curve). Label
  BO and UO indicate the bound outgoing, i.e., expanding solutions and
  unbound outgoing solutions (solid curve going on in the dotted
  curve, respectively). The horizontal lines give the range of
  observational values of the mass system.  The thin dashed curve
  separates bound and unbound regions according to the Newtonian
  criterion (above and below the thin dashed curve, respectively).}
\label{figbim}
\end{figure}

We also attempt to analyze the cluster merger between the main system
and the high velocity E-group. Even if the substructure of A2254 is
more complex, the elongation of the X-ray surface brightness suggests
that this merger is the most recent one. We apply the two-body model
(Beers et al. \cite{bee82}; Thompson \cite{tho82}) following the
methodology outlined for other DARC clusters, e.g., Abell 520 (Girardi
et al. \cite{gir08}) and Abell 2345 (Boschin et al. \cite{bos10}).
The values of the relevant parameters for the two-subclusters system
are: the above $M_{\rm sys}=1.5$--2.9 \mquii, the relative LOS
velocity in the rest-frame, $\Delta V_{\rm rf,LOS}\sim 3000$ \kss, and
the projected linear distance between the two clumps, $D\sim 0.5$ \hh,
taken as the distance between the BCG and the 2D E-peak. The
comparison between the observed shape of the X-ray surface brightness
isophotes and the simulated ones (Poole et al. \cite{poo06}) suggests
we can assume to see the merger after the core crossing and not too
much after the apocenter. For comparison with other clusters hosting
radio halos, the time $t$ elapsed from the core crossing is generally
few fractions of Gyr (e.g., Markevitch et al. \cite{mar02}; Girardi et
al. \cite{gir08}).  Figure~\ref{figbim} compares the bimodal-model
solutions as a function of $\alpha$, where $\alpha$ is the projection
angle between the plane of the sky and the line connecting the centers
of the two clumps, with the observed mass of the system $M_{\rm sys}$.
We consider two different cases ($t=0.1$ Gyr and $t=0.5$ Gyr).  In
both cases we find a bound outgoing solution (BO) with $\alpha \sim
45\degree$, and $\alpha \sim 75\degree$, respectively.  These
intermediate angles are someway in agreement with the fact that we can
detect the E-peak in the 3D analysis, which is the most efficient for
intermediate angles (Pinkney et al. \cite{pin96}). These angles lead
to a deprojected relative velocity $\Delta V_{\rm rf}\sim 3100$--4200
\ks of the subcluster with respect to the main system, comparable to
that of the bullet cluster (Markevitch et al. \cite{mar02}) or of
Abell 2744 (Boschin et al. \cite{bos06}).

$ $

In conclusion, A2254, for its large mass and unrelaxed dynamical
  status, fits well among typical clusters with radio halos described
in the literature. Present data agree with the scenario of a recent
merger with a minor subcluster, but we have evidence for a more
complex dynamical history which should be better investigated to
  explain the peculiar radio morphology.

\begin{acknowledgements}

We are in debt with Federica Govoni for the VLA radio image and with
Daniele Dallacasa, Simona Ghizzardi, and Simona Giacintucci for useful
discussions. Philip J. Humphrey is thanked for the use of his
surface-brightness fitting code and D. A. Buote is thanked for the use
of his X-ray morphology code and for useful discussions, too.  We
would like to thank Mariachiara Rossetti for the use of her
temperature map code and for useful discussions. M.G. and
F.G. acknowledge financial support from ASI-INAF (I/088/06/0 grant)
and ASI-INAF (I/009/10/0 contract), respectively. This work has been
supported by the Programa Nacional de Astronom\'\i a y Astrof\'\i sica
of the Spanish Ministry of Science and Innovation under grants
AYA2010-21322-C03-02, AYA2007-67965-C03-01 and AYA2010-21887-C04-04.
This publication is based on observations made on the island of La
Palma with the Italian Telescopio Nazionale Galileo (TNG) and the
Isaac Newton Telescope (INT). The TNG is operated by the Fundaci\'on
Galileo Galilei -- INAF (Istituto Nazionale di Astrofisica). The INT
is operated by the Isaac Newton Group. Both telescopes are located in
the Spanish Observatorio of the Roque de Los Muchachos of the
Instituto de Astrofisica de Canarias.  This publication is also based
on observations obtained with XMM-Newton, an ESA science mission with
instruments and contributions directly funded by ESA Member States and
NASA.  This research has made use of the NASA/IPAC Extragalactic
Database (NED), which is operated by the Jet Propulsion Laboratory,
California Institute of Technology, under contract with the National
Aeronautics and Space Administration.

\end{acknowledgements}

\end{document}

%% file: catalogA2254.tex

\begin{table}[!ht]
        \caption[]{Velocity catalog of 128 spectroscopically measured
          galaxies in the field of the cluster A2254. }
         \label{catalogA2254}
              $$ 
           \begin{array}{r r c c c c r r}
            \hline
            \noalign{\smallskip}
            \hline
            \noalign{\smallskip}

\mathrm{ID} & \mathrm{IDm} &\mathrm{\alpha},\mathrm{\delta}\,(\mathrm{J}2000)  & g^{\prime} & r^{\prime}& i^{\prime}& \mathrm{v}\,\,\,\,\,&\mathrm{\Delta}\mathrm{v}\\
  & & & & & &\mathrm{(\,km}&\mathrm{s^{-1}\,)}\\
            \hline
            \noalign{\smallskip}  

  1           &  1           &17\ 17\ 24.62 ,+19\ 41\ 59.6&   18.52&     17.45& 16.98&   51510&  53  \\
  2           &  2           &17\ 17\ 25.92 ,+19\ 39\ 54.7&   20.72&     19.63& 19.25&   52688&  68  \\ 
        3     &  -          &17\ 17\ 25.97 ,+19\ 43\ 25.3&   18.62&     17.27& 16.84&   72033&  64  \\ 
        4     &  -    &17\ 17\ 26.26 ,+19\ 42\ 20.5&   18.90&     17.70& 17.41&   70536&  66  \\ 
  5           &  3            &17\ 17\ 26.59 ,+19\ 41\ 31.6&   18.89&     18.17& 17.96&   52741&  77  \\ 
  6           &  4           &17\ 17\ 27.62 ,+19\ 42\ 14.8&   19.31&     18.15& 17.74&   52931&  46  \\ 
        7     &  -    &17\ 17\ 28.94 ,+19\ 40\ 18.1&   21.59&     20.08& 19.37&  165420& 106  \\ 
  8           &  5           &17\ 17\ 30.10 ,+19\ 43\ 30.0&   19.88&     18.71& 18.30&   53232&  70  \\ 
  9           &  6           &17\ 17\ 31.68 ,+19\ 42\ 06.1&   19.28&     18.16& 17.76&   53803&  55  \\ 
        10    &  -   &17\ 17\ 32.35 ,+19\ 43\ 13.4&   20.82&     19.89& 19.46&   78076&  84  \\ 
 11           & 7          &17\ 17\ 33.53 ,+19\ 41\ 07.1&   19.68&     19.11& 18.99&   53136&  90  \\ 
        12    & -    &17\ 17\ 33.55 ,+19\ 42\ 42.1&   20.26&     19.21& 18.94&   71797&  64  \\ 
        13    & -  &17\ 17\ 34.37 ,+19\ 42\ 48.6&   17.50&     16.51& 16.12&   36622&  37  \\ 
 14           &  8           &17\ 17\ 35.35 ,+19\ 42\ 42.8&   19.70&     18.87& 18.56&   54559& 103  \\ 
 15           &  9           &17\ 17\ 35.95 ,+19\ 39\ 27.4&   20.10&     19.10& 18.81&   52006& 139  \\ 
 16           & 10           &17\ 17\ 36.43 ,+19\ 41\ 07.8&   19.71&     18.67& 18.32&   51360&  42  \\ 
 17           & 11           &17\ 17\ 36.46 ,+19\ 39\ 02.2&   19.32&     18.57& 18.26&   54600&  73  \\ 
 18           & 12           &17\ 17\ 36.74 ,+19\ 42\ 03.6&   20.80&     19.65& 19.28&   54105&  70  \\ 
 19           & 13           &17\ 17\ 37.03 ,+19\ 41\ 11.4&   19.44&     18.75& 18.35&   53377&  53  \\ 
         20   & -   &17\ 17\ 37.87 ,+19\ 41\ 17.9&   21.04&     19.72& 19.36&   62831&  62  \\ 
         21   & -   &17\ 17\ 38.04 ,+19\ 40\ 04.4&   20.20&     18.95& 18.72&   70432&  46  \\ 
 22           & 14           &17\ 17\ 38.04 ,+19\ 44\ 55.7&   20.37&     19.27& 18.89&   55154& 101  \\ 
 23           & 15           &17\ 17\ 38.21 ,+19\ 40\ 03.7&   21.76&     20.84& 19.88&   52576& 150  \\ 
 24           & 16           &17\ 17\ 38.26 ,+19\ 38\ 46.3&   21.17&     20.03& 19.47&   53964& 100  \\ 
 25           & 17           &17\ 17\ 38.95 ,+19\ 40\ 01.6&   19.37&     18.20& 17.79&   52664&  62  \\ 
         26   & -   &17\ 17\ 39.10 ,+19\ 41\ 33.0&   21.20&     19.77& 19.21&  149288&  48  \\ 
 27           & 18           &17\ 17\ 39.65 ,+19\ 41\ 06.7&   20.22&     19.05& 18.79&   51500& 121  \\ 
 28           & 19           &17\ 17\ 39.70 ,+19\ 38\ 02.8&   20.52&     19.45& 19.07&   54068&  63  \\ 
 29           & 20           &17\ 17\ 39.84 ,+19\ 38\ 10.7&   20.67&     19.64& 19.30&   52684&  90  \\ 
 30           & 21           &17\ 17\ 39.84 ,+19\ 38\ 10.7&   20.67&     19.64& 19.30&   54818&  90  \\ 
 31           & 22           &17\ 17\ 40.03 ,+19\ 40\ 52.0&   19.43&     18.35& 17.98&   52821&  33  \\ 
 32           & 23           &17\ 17\ 40.63 ,+19\ 37\ 42.6&   20.52&     19.86& 19.63&   51947& 240  \\ 
 33           & 24           &17\ 17\ 40.66 ,+19\ 41\ 40.9&   20.33&     19.16& 18.70&   54364&  66  \\ 
 34           & 25           &17\ 17\ 40.68 ,+19\ 39\ 52.2&   20.98&     19.84& 19.39&   54308& 103  \\ 
 35           & 26           &17\ 17\ 41.02 ,+19\ 43\ 53.8&   20.16&     19.37& 19.06&   51559& 100  \\ 
 36           & 27           &17\ 17\ 41.33 ,+19\ 39\ 52.2&   19.78&     19.06& 18.81&   53081& 128  \\ 
         37   & -   &17\ 17\ 41.38 ,+19\ 41\ 42.0&   18.73&     18.04& 17.75&   62608&  55  \\ 
 38           & 28           &17\ 17\ 41.57 ,+19\ 41\ 07.4&   19.01&     18.04& 17.81&   54845&  66  \\ 
 39           & 29           &17\ 17\ 42.34 ,+19\ 42\ 39.2&   17.98&     17.18& 16.90&   54521&  77  \\ 
 40           & 30           &17\ 17\ 43.01 ,+19\ 42\ 26.3&   19.19&     18.05& 17.58&   53870&  62  \\ 
 41           & 31           &17\ 17\ 43.06 ,+19\ 41\ 44.2&   18.91&     17.98& 17.57&   52292&  35  \\ 
 42           & 32           &17\ 17\ 43.06 ,+19\ 43\ 51.6&   18.75&     17.69& 17.41&   53468&  41  \\   
                              
                                       \noalign{\smallskip}			    
            \hline					    
            \noalign{\smallskip}			    
            \hline					    
         \end{array}
     $$        
         \end{table}
\addtocounter{table}{-1}
\begin{table}[!ht]
          \caption[ ]{Continued.}
     $$        
           \begin{array}{r r c c c c r r}
            \hline
            \noalign{\smallskip}
            \hline
            \noalign{\smallskip}
               
\mathrm{ID} & \mathrm{IDm} &\mathrm{\alpha},\mathrm{\delta}\,(\mathrm{J}2000)  & g^{\prime} & r^{\prime}& i^{\prime}& \mathrm{v}\,\,\,\,\,&\mathrm{\Delta}\mathrm{v}\\
  & & & & & &\mathrm{(\,km}&\mathrm{s^{-1}\,)}\\
               
            \hline
            \noalign{\smallskip}
                              
 43          & 33          &17\ 17\ 43.20 ,+19\ 44\ 59.6&   19.16&     18.28& 17.99&   50532&  51  \\
 44          & 34          &17\ 17\ 43.61 ,+19\ 39\ 20.5&   19.25&     18.10& 17.77&   52880&  66  \\
 45          & 35          &17\ 17\ 43.61 ,+19\ 42\ 42.1&   19.58&     18.47& 18.09&   53563&  44  \\
 46          & 36          &17\ 17\ 43.63 ,+19\ 38\ 37.7&   21.34&     20.46& 20.07&   51447&  99  \\
 47          & 37          &17\ 17\ 43.70 ,+19\ 39\ 24.8&   20.90&     19.97& 19.65&   52533& 189  \\
 48          & 38          &17\ 17\ 43.80 ,+19\ 41\ 12.1&   19.26&     18.24& 17.96&   55744&  62  \\
 49          & 39          &17\ 17\ 43.87 ,+19\ 41\ 43.4&   19.42&     18.10& 17.93&   52024&  66  \\
 50          & 40          &17\ 17\ 45.53 ,+19\ 40\ 40.1&   17.90&     17.01& 16.74&   54579&  53  \\
 51          & 41          &17\ 17\ 45.62 ,+19\ 41\ 57.1&   21.28&     20.09& 19.72&   53513& 110  \\
 52          & 42          &17\ 17\ 45.65 ,+19\ 38\ 13.6&   19.31&     18.33& 17.93&   53918&  66  \\
 53          & 43          &17\ 17\ 45.86 ,+19\ 40\ 48.4&   17.44&     16.22& 15.73&   52963&  41  \\
 54          & 44          &17\ 17\ 45.86 ,+19\ 41\ 47.4&   19.69&     18.62& 18.23&   51347&  55  \\
 55          & 45          &17\ 17\ 45.86 ,+19\ 40\ 45.8&   20.66&     19.76& 19.21&   53276&  75  \\
 56          & 46          &17\ 17\ 45.91 ,+19\ 39\ 41.4&   19.45&     18.29& 17.89&   53264&  44  \\
 57          & 47          &17\ 17\ 46.30 ,+19\ 42\ 20.9&   20.35&     19.24& 18.84&   52108& 130  \\
 58          & 48          &17\ 17\ 46.51 ,+19\ 42\ 42.5&   20.45&     19.49& 19.01&   53167&  59  \\
 59          & 49          &17\ 17\ 46.58 ,+19\ 40\ 05.9&   19.90&     19.27& 18.40&   53069&  55  \\
        60   & -  &17\ 17\ 46.75 ,+19\ 40\ 30.7&   19.43&     18.83& 18.58&   23305& 170  \\
 61          & 50          &17\ 17\ 46.97 ,+19\ 39\ 31.3&   18.94&     17.81& 17.36&   50067&  66  \\
        62   & -  &17\ 17\ 47.59 ,+19\ 41\ 29.4&   20.50&     19.78& 19.51&   89174& 158  \\
 63          & 51          &17\ 17\ 47.69 ,+19\ 44\ 15.4&   19.27&     18.46& 18.20&   55316&  66  \\
 64          & 52          &17\ 17\ 47.78 ,+19\ 41\ 35.9&   19.13&     18.14& 17.80&   54202&  59  \\
 65          & 53          &17\ 17\ 47.95 ,+19\ 37\ 00.8&   20.57&     19.49& 19.07&   55387&  59  \\
 66          & 54          &17\ 17\ 48.02 ,+19\ 38\ 51.4&   20.31&     19.28& 18.92&   55179&  46  \\
 67          & 55          &17\ 17\ 49.78 ,+19\ 41\ 10.7&   19.82&     18.87& 18.55&   54240&  77  \\
 68          & 56          &17\ 17\ 49.82 ,+19\ 41\ 18.2&   21.19&     20.17& 19.85&   51305& 114  \\
 69          & 57          &17\ 17\ 50.02 ,+19\ 39\ 07.6&   18.15&     16.98& 16.57&   54695&  39  \\
 70          & 58          &17\ 17\ 50.81 ,+19\ 41\ 43.4&   18.86&     17.76& 17.35&   51441&  37  \\
 71          & 59          &17\ 17\ 50.95 ,+19\ 41\ 12.5&   20.47&     19.41& 19.28&   53289&  55  \\
 72          & 60          &17\ 17\ 51.22 ,+19\ 42\ 38.5&   18.84&     18.06& 17.84&   49237&  79  \\
 73          & 61          &17\ 17\ 51.48 ,+19\ 40\ 36.1&   19.47&     18.31& 17.95&   52695&  79  \\
        74   &  -  &17\ 17\ 51.58 ,+19\ 42\ 52.9&   18.61&     17.37& 16.94&   62934&  70  \\
 75          & 62          &17\ 17\ 51.65 ,+19\ 41\ 20.4&   19.78&     18.57& 18.70&   51961&  86  \\
 76          & 63          &17\ 17\ 52.13 ,+19\ 40\ 19.6&   19.39&     18.26& 17.85&   52167&  59  \\
 77          & 64          &17\ 17\ 52.61 ,+19\ 40\ 08.0&   20.18&     19.04& 18.76&   52308&  46  \\
 78          & 65          &17\ 17\ 53.04 ,+19\ 44\ 24.0&   19.94&     18.82& 18.45&   51433&  66  \\
 79          & 66          &17\ 17\ 53.21 ,+19\ 38\ 47.8&   18.73&     17.68& 17.34&   50868&  46  \\
 80          & 67          &17\ 17\ 53.52 ,+19\ 40\ 25.0&   20.20&     19.10& 18.68&   56052&  86  \\
 81          & 68          &17\ 17\ 54.12 ,+19\ 41\ 47.8&   19.13&     18.00& 17.64&   52069&  48  \\
 82          & 69          &17\ 17\ 54.17 ,+19\ 43\ 36.1&   21.78&     20.77& 20.30&   53213& 108  \\
 83          & 70          &17\ 17\ 54.84 ,+19\ 38\ 14.3&   19.58&     18.70& 18.37&   53855&  64  \\
 84          & 71          &17\ 17\ 54.91 ,+19\ 39\ 04.3&   20.28&     19.16& 18.74&   53449&  75  \\        
                              
                                       \noalign{\smallskip}			    
            \hline					    
            \noalign{\smallskip}			    
            \hline					    
         \end{array}
     $$        
         \end{table}
\addtocounter{table}{-1}
\begin{table}[!ht]
          \caption[ ]{Continued.}
     $$        
           \begin{array}{r r c c c c r r}
            \hline
            \noalign{\smallskip}
            \hline
            \noalign{\smallskip}
               
\mathrm{ID} & \mathrm{IDm} &\mathrm{\alpha},\mathrm{\delta}\,(\mathrm{J}2000)  & g^{\prime} & r^{\prime}& i^{\prime}& \mathrm{v}\,\,\,\,\,&\mathrm{\Delta}\mathrm{v}\\
  & & & & & &\mathrm{(\,km}&\mathrm{s^{-1}\,)}\\

            \hline
            \noalign{\smallskip}
                              
 85           & 72           &17\ 17\ 54.96 ,+19\ 39\ 29.2&   20.38&     19.87& 19.61&   54529& 174  \\  
 86           & 73           &17\ 17\ 55.06 ,+19\ 40\ 16.0&   18.56&     17.49& 17.11&   49834&  62  \\  
 87           & 74           &17\ 17\ 55.06 ,+19\ 42\ 10.8&   20.40&     19.22& 18.82&   56376&  66  \\  
 88           & 75           &17\ 17\ 55.08 ,+19\ 41\ 10.7&   19.59&     18.40& 17.99&   56400&  48  \\  
 89           & 76           &17\ 17\ 55.46 ,+19\ 37\ 52.0&   21.36&     20.34& 19.98&   53505& 101  \\  
 90           & 77           &17\ 17\ 55.75 ,+19\ 41\ 53.2&   20.84&     19.94& 19.64&   56368&  97  \\  
 91           & 78           &17\ 17\ 55.78 ,+19\ 39\ 41.4&   19.97&     18.83& 18.44&   56515&  59  \\  
 92           & 79           &17\ 17\ 55.78 ,+19\ 40\ 36.8&   19.18&     18.60& 18.37&   52917&  68  \\  
 93           & 80           &17\ 17\ 56.02 ,+19\ 40\ 59.5&   18.93&     17.81& 17.41&   53106&  53  \\  
 94           & 81           &17\ 17\ 56.09 ,+19\ 42\ 14.4&   20.78&     19.93& 19.98&   51396&  75  \\  
 95           & 82           &17\ 17\ 56.30 ,+19\ 39\ 54.4&   19.64&     18.49& 18.05&   51354&  37  \\  
 96           & 83           &17\ 17\ 56.42 ,+19\ 41\ 43.4&   20.13&     19.15& 18.82&   55943& 117  \\  
 97           & 84           &17\ 17\ 56.50 ,+19\ 37\ 59.9&   20.34&     19.25& 18.86&   53815&  57  \\  
 98           & 85           &17\ 17\ 56.62 ,+19\ 42\ 56.5&   19.35&     18.18& 17.82&   54648&  35  \\  
 99           & 86           &17\ 17\ 56.95 ,+19\ 41\ 21.5&   19.97&     18.92& 18.59&   54154&  44  \\  
100           & 87           &17\ 17\ 57.07 ,+19\ 41\ 28.7&   18.67&     17.54& 17.16&   53106&  42  \\  
101           & 88           &17\ 17\ 57.19 ,+19\ 37\ 23.5&   20.86&     19.86& 19.61&   52885&  97  \\  
102           & 89           &17\ 17\ 57.26 ,+19\ 39\ 35.3&   17.79&     16.66& 16.25&   51409&  55  \\
103           & 90           &17\ 17\ 57.29 ,+19\ 41\ 52.4&   18.92&     17.78& 17.38&   52138&  42  \\  
104           & 91           &17\ 17\ 57.53 ,+19\ 43\ 52.0&   18.91&     17.75& 17.59&   53675&  51  \\  
105           & 92           &17\ 17\ 57.67 ,+19\ 43\ 42.2&   19.98&     19.06& 19.05&   53747&  79  \\  
106           & 93           &17\ 17\ 57.84 ,+19\ 43\ 28.9&   19.01&     18.45& 18.17&   54201&  68  \\  
107           & 94           &17\ 17\ 58.13 ,+19\ 38\ 39.1&   18.47&     17.59& 17.41&   50570&  55  \\  
108           & 95           &17\ 17\ 58.20 ,+19\ 36\ 45.7&   20.65&     19.54& 19.23&   51708&  62  \\  
        109   &  -  &17\ 17\ 58.20 ,+19\ 40\ 36.5&   20.34&     19.60& 19.20&   96562& 147  \\  
110           & 96           &17\ 17\ 59.21 ,+19\ 37\ 23.2&   19.97&     18.93& 18.59&   51871&  70  \\  
111           & 97           &17\ 17\ 59.21 ,+19\ 40\ 45.1&   19.34&     18.19& 17.73&   56757&  55  \\  
112           & 98           &17\ 17\ 59.69 ,+19\ 40\ 25.7&   17.77&     16.67& 16.26&   53416&  64  \\
113           & 99           &17\ 17\ 59.83 ,+19\ 42\ 09.0&   19.57&     18.92& 18.70&   51016&  57  \\  
114           &100           &17\ 17\ 59.98 ,+19\ 40\ 54.1&   20.66&     19.89& 19.57&   53090&  68  \\  
115           &101           &17\ 18\ 00.58 ,+19\ 43\ 44.8&   20.94&     20.04& 19.64&   52104&  90  \\  
116           &102           &17\ 18\ 00.67 ,+19\ 40\ 00.1&   20.92&     19.80& 19.40&   54904&  77  \\  
117           &103           &17\ 18\ 01.30 ,+19\ 40\ 58.4&   17.74&     16.64& 16.21&   54229&  57  \\
118           &104           &17\ 18\ 01.70 ,+19\ 39\ 28.8&   19.27&     18.13& 17.72&   52903&  68  \\  
119           &105           &17\ 18\ 01.97 ,+19\ 40\ 22.4&   19.56&     18.53& 18.08&   51933&  92  \\  
        120   &  -  &17\ 18\ 01.99 ,+19\ 37\ 56.6&   20.09&     19.31& 18.97&   70025&  99  \\  
121           &106           &17\ 18\ 03.31 ,+19\ 42\ 09.4&   19.80&     18.70& 18.34&   55067&  46  \\  
122           &107           &17\ 18\ 03.46 ,+19\ 36\ 14.0&   20.19&     19.12& 18.73&   51813&  64  \\  
        123   &   -  &17\ 18\ 03.62 ,+19\ 37\ 33.6&   21.07&     19.97& 19.51&  122887& 200  \\  
124           &108           &17\ 18\ 04.46 ,+19\ 40\ 21.4&   19.72&     19.58& 19.50&   50836& 110  \\  
        125   &   -  &17\ 18\ 04.61 ,+19\ 39\ 22.3&   19.71&     18.28& 17.80&  105488&  55  \\  
126           &109           &17\ 18\ 06.00 ,+19\ 39\ 55.1&   19.51&     18.52& 18.17&   52452& 125  \\  
127           &110           &17\ 18\ 06.05 ,+19\ 38\ 17.5&   18.38&     17.29& 16.90&   51558&  62  \\  
        128   &   -  &17\ 18\ 07.51 ,+19\ 37\ 40.8&   19.72&     18.71& 18.36&   96285&  75  \\        

                        \noalign{\smallskip}			    
            \hline					    
            \noalign{\smallskip}			    
            \hline					    
         \end{array}
     $$ 
\end{table}


%% file: tabdedica2d.tex
\begin{table}
        \caption[]{2D substructure from the INT and SUBARU photometric samples.}
         \label{tabdedica2d}
            $$
         \begin{array}{l r c c c }
            \hline
            \noalign{\smallskip}
            \hline
            \noalign{\smallskip}
\mathrm{Subclump} & N_{\rm S} & \alpha({\rm J}2000),\,\delta({\rm J}2000)&\rho_{
\rm S}&\chi^2_{\rm S}\\
& & \mathrm{h:m:s,\degree:\arcmm:\arcs}&&\\
         \hline
         \noalign{\smallskip}
\mathrm{2D-E\ (INT\ r'<21)}    &164&17\ 17\ 55.2+19\ 41\ 13&1.00&65\\
\mathrm{2D-W\ (INT\ r'<21)}    &214&17\ 17\ 48.0+19\ 41\ 44&0.91&55\\
\mathrm{2D-ext\ (INT\ r'<21)}  &159&17\ 16\ 31.4+19\ 34\ 48&0.33&44\\
\mathrm{2D-E\ (SUB\ i'<20.5)}    &131&17\ 17\ 55.9+19\ 41\ 03&1.00&43\\
\mathrm{2D-W\ (SUB\ i'<20.5)}    &248&17\ 17\ 46.8+19\ 41\ 07&0.94&46\\
\mathrm{2D-ext\ (SUB\ i'<20.5)}  &159&17\ 16\ 38.0+19\ 35\ 33&0.37&33\\
\mathrm{2D-SW\ (SUB\ i'<21.5)}  &122&17\ 17\ 47.0+19\ 40\ 53&1.00&37\\
\mathrm{2D-E\ (SUB\ i'<21.5)} & 77&17\ 17\ 56.4+19\ 41\ 13&0.73&31\\
\mathrm{2D-NW\ (SUB\ i'<21.5)} & 22&17\ 17\ 46.0+19\ 42\ 10&0.69&15\\
\mathrm{2D-ext\ (SUB\ i'<21.5)}& 54&17\ 16\ 36.5+19\ 35\ 45&0.36&19\\
              \noalign{\smallskip}
              \noalign{\smallskip}
            \hline
            \noalign{\smallskip}
            \hline
         \end{array}
$$
         \end{table}

%% file: tabsub.tex
\begin{table*}
        \caption[]{Kinematical properties of the whole system and 
galaxy subsystems.}
         \label{tabsub}
            $$
         \begin{array}{l r r r c  r r r}
            \hline
            \noalign{\smallskip}
\mathrm{Substructure\ results}&&&&&\mathrm{3D\ KMM}&&\\ 
            \hline
            \noalign{\smallskip}
\mathrm{System} & N_{\rm g} &\mathrm{<v>}&\sigma_{\rm V}&\mathrm{Interpretation}& N_{\rm g,KMM} &\mathrm{<v>}_{KMM}&\sigma_{\rm v,KMM}\\
& &\mathrm{km\ s^{-1}}&\mathrm{km\ s^{-1}}&&&\mathrm{km\ s^{-1}}&\mathrm{km\ s^{-1}}\\
         \hline
         \noalign{\smallskip}
\mathrm{Whole\ system} &110& 53128\pm128& 1340_{\ 84}^{101} &\mathrm{-}&\mathrm{-}&\mathrm{-}&\mathrm{-}\\
\mathrm{Eastern\ system} & 58& 53098\pm208& 1576_{104}^{168}&\mathrm{contaminated\ main}&\mathrm{-}&\mathrm{-}&\mathrm{-}\\
\mathrm{Western\ system}& 52& 53186\pm150& 1073_{\ 92}^{118}&\mathrm{uncontaminated\ main}&\mathrm{-}&\mathrm{-}&\mathrm{-}\\
\mathrm{E1.5HV}        &  6& 56310\pm226&  245_{180}^{\ 86} &\mathrm{high\ vel.\ E-group}&  7& 56414\pm116&  370_{181}^{291}  \\
\mathrm{E1.5LV}        & 21& 52463\pm341& 1000_{102}^{220}  &\mathrm{main}&103& 52968\pm119& 1180_{\ 67}^{\ 88}\\
\mathrm{HT2}           &  7& 56355\pm244&  232_{104}^{\ 68} &\mathrm{high\ vel.\ E-group}&  8& 56341\pm115&  469_{255}^{183} \\ 
\mathrm{HT1}           & 96& 52980\pm108& 1053_{\ 66}^{\ 59}&\mathrm{(substructured)\ main}&\mathrm{-}&\mathrm{-}&\mathrm{-}\\
\mathrm{HT12}          &  8& 51414\pm159&  421_{\ 91}^{127} &\mathrm{low\ vel.\ W-main}& 35& 51840\pm100&  889_{\ 63}^{162}\\ 
\mathrm{HT11}          & 42& 53299\pm188&  739_{\ 55}^{\ 81}&\mathrm{principal\ main}& 67& 53550\pm153&  939_{\ 69}^{\ 86}\\ 
\mathrm{DED1}          &  3& 56322\pm139&  197^{\mathrm{a}\ \ }_{\ \ \ }&\mathrm{high\ vel.\ E-group}&  4& 56422\pm109&  167_{132}^{\ 70}\\ 
\mathrm{DED2}          &  9& 54313\pm213&  579_{\ 80}^{118}&\mathrm{NW-group}&  27& 54246\pm130&  663_{\ 68}^{\ 98}\\ 
\mathrm{DED3}          & 12& 53151\pm362& 1171_{134}^{248}&\mathrm{E\ group}&   26& 53385\pm320& 1591_{193}^{329}\\ 
\mathrm{DED4}          & 30& 53054\pm177&  950_{\ 98}^{141}&\mathrm{SW\ group}&  46& 52168\pm114&  763_{\ 66}^{\ 84}\\ 
\mathrm{DED5}          &  4& 51798\pm293&  448_{\ 46}^{\ 21}&\mathrm{low\ vel.\ SE-group}&  7& 53238\pm429& 1000_{\ 59}^{407}\\ 
              \noalign{\smallskip}
             \noalign{\smallskip}
            \hline
            \noalign{\smallskip}
            \hline
         \end{array}
$$
\begin{list}{}{}  
\item[$^{\mathrm{a}}$] Standard velocity dispersion estimate.
\end{list}
         \end{table*}

%% file: tabxmorpho.tex
\begin{table}
        \caption[]{Results of the morphological analysis of the X-ray
          image of A2254 using power ratios, centroid shifts and
          concentration.}
         \label{tabxmorpho}
            $$
         \begin{array}{c c c c c c}
            \hline
            \noalign{\smallskip}
            \hline
            \noalign{\smallskip}
\mathrm{P^{pk}_1/P^{pk}_0} & \mathrm{P_2/P_0} &\mathrm{P_3/P_0} & \mathrm{P_4/P_0} &w & c \\
(\times 10^{-4}) & (\times 10^{-7}) & (\times 10^{-7}) & (\times 10^{-8})& (500\,\mathrm{h_{70}^{-1}\,kpc})& \\
         \hline
         \noalign{\smallskip}
1.84\pm0.47 & 4.25\pm1.26&2.51\pm0.82&1.29\pm1.12& 0.037\pm0.001 & 0.08 \\
              \noalign{\smallskip}
             \noalign{\smallskip}
            \hline
            \noalign{\smallskip}
            \hline
         \end{array}
$$
         \end{table}